\documentclass[]{spie}  

\usepackage{amsmath,amsfonts,amssymb}
\usepackage{graphicx}
\usepackage[colorlinks=true, allcolors=blue]{hyperref}

\title{Astrophotonics: photonic integrated circuits for astronomical instrumentation}

\author[a,b]{Martin M. Roth}
\author[a]{Kalaga Madhav}
\author[a,b]{Andreas Stoll}
\author[a,b]{Daniel Bodenm\"uller}
\author[a]{Aline Dinkelaker}
\author[a]{Aashia Rahman}
\author[a]{Eloy Hernandez}
\author[a]{Alan G\"unther}
\author[a]{Stella Vjesnica}

\affil[a]{Leibniz-Institute for Astrophysics Potsdam (AIP), An der Sternwarte 16, 14482 Potsdam, Germany}
\affil[b]{Universit\"at Potsdam, Institut für Physik und Astronomie, Haus 28, Karl-Liebknecht-Straße 24/25, 14476 Potsdam, Germany}

\authorinfo{Further author information: mmroth@aip.de}

\begin{document} 
\maketitle

\begin{abstract}
Photonic Integrated Circuits (PIC) are best known for their important role in the telecommunication sector, e.g. high speed communication devices in data centers. However, PIC also hold the promise for innovation in sectors like life science, medicine, sensing, automotive etc. The past two decades have seen efforts of utilizing PIC to enhance the performance of instrumentation for astronomical telescopes, perhaps the most spectacular example being the integrated optics beam combiner for
the interferometer GRAVITY at the ESO Very Large Telescope. This instrument has enabled observations of the supermassive black hole in the center of the Milky Way at unprecedented angular resolution, eventually leading to the Nobel Price for Physics in 2020. Several groups worldwide are actively engaged in the emerging field of astrophotonics research, amongst them the innoFSPEC Center in Potsdam, Germany. We present results for a number of applications developed at innoFSPEC, notably PIC for integrated photonic spectrographs on the basis of arrayed waveguide gratings and the PAWS demonstrator (Potsdam Arrayed Waveguide Spectrograph), PIC-based ring resonators in astronomical frequency combs for precision wavelength calibration, discrete beam combiners (DBC) for large astronomical interferometers, as well as aperiodic fiber Bragg gratings for complex astronomical filters and their possible derivatives
in PIC.
\end{abstract}


\section{INTRODUCTION}
\label{sec:intro}  
Since the first use of lenses for a telescope by Galileo Galilei [\citenum{Galilei1610}] back in 1610, astronomical instrumentation has been key for new discoveries, and shaping our picture of the universe. In the context of photonics, it is interesting to note that early on, from the first availability of optical fibers, astronomers have discovered their virtue for spectroscopy: not only do fibers provide a higher level of flexibility than free space optics in coupling spectrographs at practically arbitrary locations to the telescope focal plane [\citenum{Angel1977,Angel1979}], but also an elegant way to accomplish multi object spectroscopy (MOS) [\citenum{Hill1980}], i.e. the opportunity to measure spectra of N objects at the same time, rather than sequentially. Given the necessary total exposure times for faint objects, that often extend to hours rather than only seconds or minutes, the need to create databases of hundreds of thousands, or even millions, of objects has made fiber coupled MOS an industry, enabling e.g. fundamental studies of the large scale structure of the universe to validate numerical simulations in cosmology. For example, the Sloan Digital Sky Survey (SDSS) [\citenum{Gunn2006}], arguably one of the highest scientific impact ground based observing programs ever, has employed fiber-coupled MOS to provide hundreds of thousands of spectra for stars and galaxies in a series of data releases, spanning a period of more than 20 years. High resolution echelle spectrographs with thermal and pressure control have achieved unprecedented precision in measuring Doppler shifts of stars for exoplanet detection, thanks to a flexure-free stationary setup that is coupled to the telescope by means of a fiber link, e.g. [\citenum{Mayor2003}], [\citenum{Strassmeier2015}], [\citenum{Pepe2021}]. Also, fiber bundles have enabled spectroscopy over an extended two-dimensional field-of-view, e.g. [\citenum{Roth2005}], [\citenum{Kelz2006}], with powerful astronomical applications, e.g. the study of the evolution of galaxies [\citenum{Sanchez2012}], [\citenum{Bundy2015}].

Beyond the application of optical fibers, the last two decades have seen more sophisticated applications of waveguides for astronomical instrumentation, such as to coin the term ''Astrophotonics'' [\citenum{Bland-Hawthorn2004}]. The emerging field of Astrophotonics is sketched in [\citenum{Norris2019}], and more comprehensively described in a recent review [\citenum{Minardi2021}]. Here we summarize results achieved at the innovation center innoFSPEC Potsdam [\citenum{Roth2008}], [\citenum{Haynes2010}] [\citenum{Gatkine2019}] with a focus on photonic integrated circuits.

\section{Integrated photonic spectrographs on the basis of arrayed waveguide gratings}
\label{sec:AWG} 

Fiber-coupled miniaturized spectrometers based on classical diffraction gratings have been available as commercial products for some time, but these devices are generally not suitable for demanding applications in astronomical instrumentation. However, the demand for dense wavelength domain multiplexing (DWDM) in telecommunications and the development of arrayed waveguide gratings (AWG) has inspired the transfer of this technologies to other fields, such as spectroscopy in medicine, life sciences, and chemistry, e.g. [\citenum{Cheben2007}]. First experiments into this direction for astronomy [\citenum{Bland-Hawthorn2006}] were reported by [\citenum{Cvetojevic2009}] and [\citenum{Cvetojevic2012}], and a vision for future development proposed by [\citenum{Bland-Hawthorn2010}]. The early experiments were able to demonstrate a proof of principle by using commercially available AWGs, however falling short of delivering a performance needed for a competitive focal plane instrument at an observatory. 

The innoFSPEC innovation center has engaged in research for integrated photonic spectrographs that are optimized for astronomy. For the time being, the focus has remained on a wavelength region in the near infrared (NIR), namely the astronomical H-band (1500 $\ldots$ 1800~nm), that is close enough to telecommunication wavelengths to facilitate manufacture and characterization.

\begin{figure} [h]
   \begin{center}
    \begin{tabular}{c} 
   \includegraphics[height=65mm]{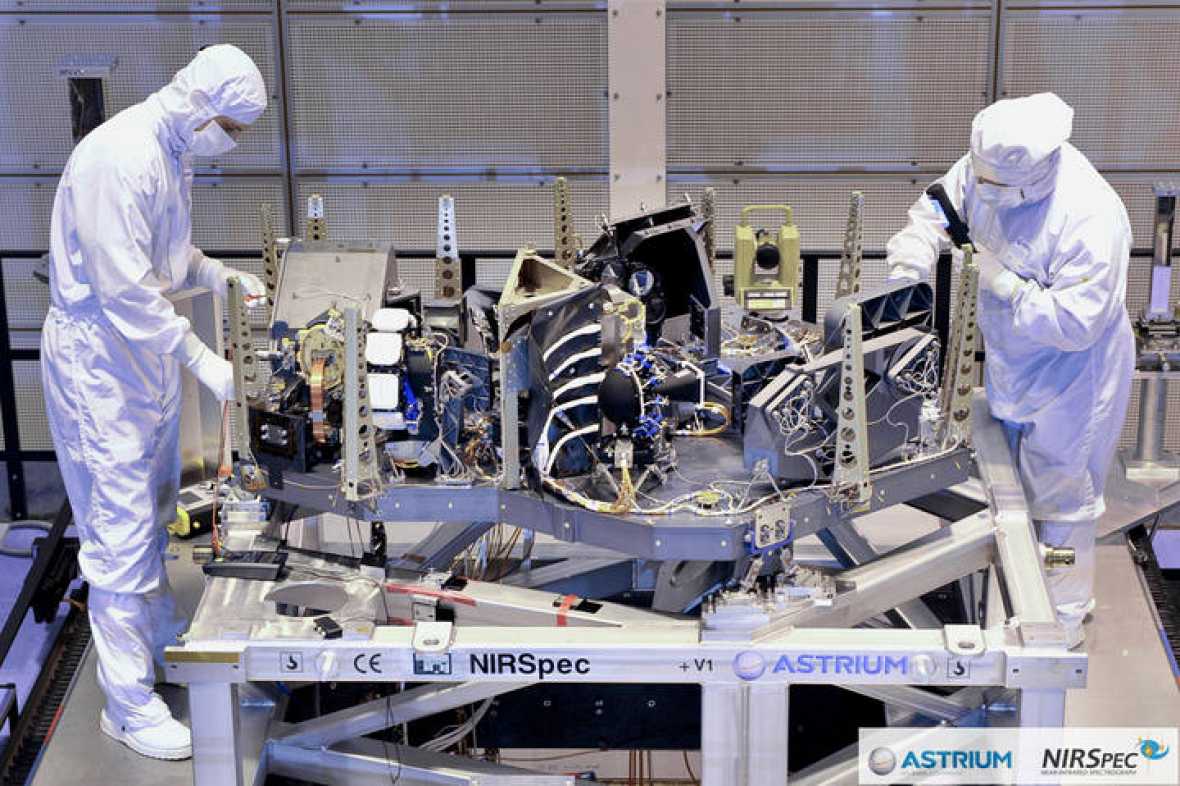}
   \hspace{3mm}
   \includegraphics[height=65mm]{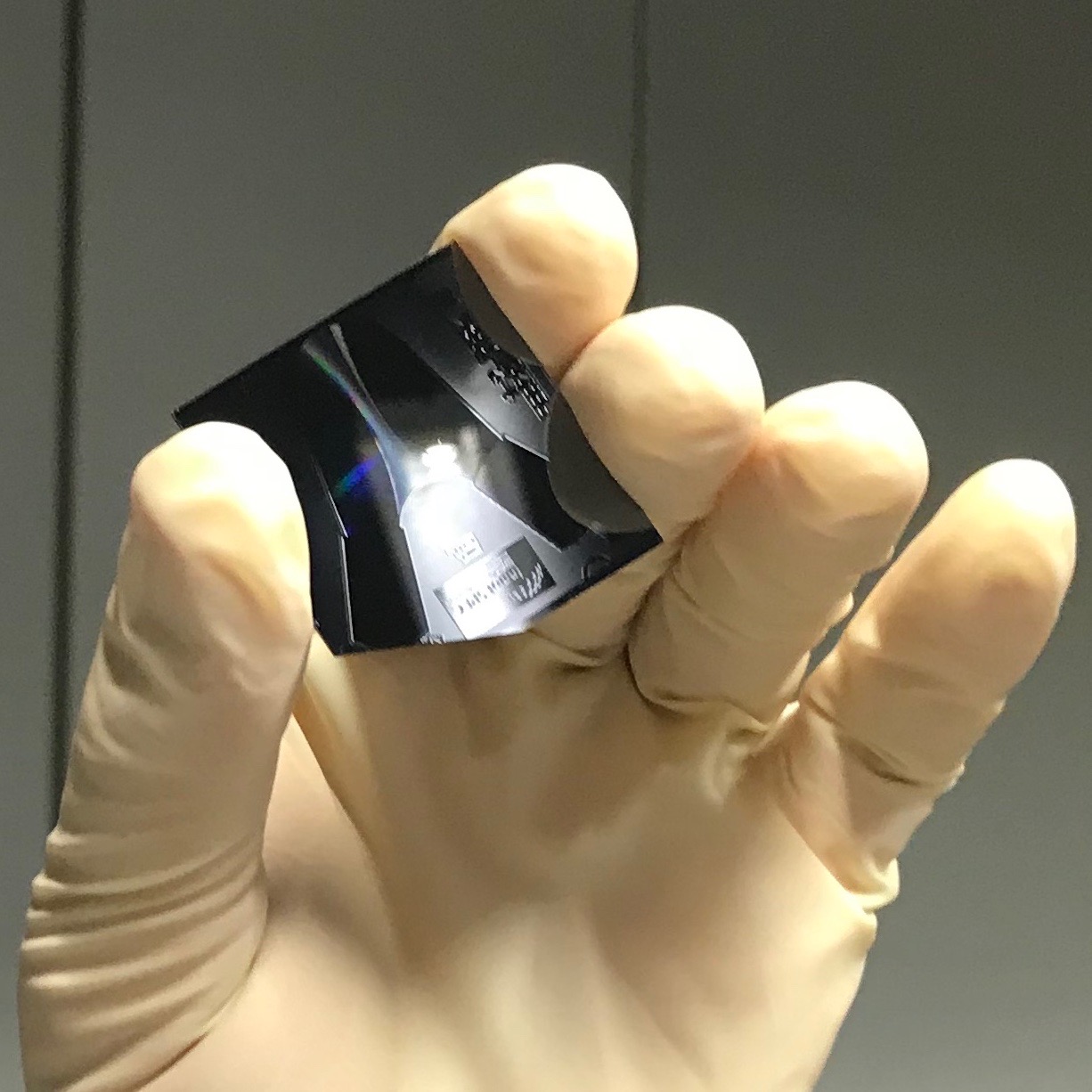}
	\end{tabular}
	\end{center}
   \caption[example] 
   { \label{fig:1} 
Miniaturizing a free-space optical system spectrograph to a chip. Left: NIRspec instrument on board of JWST (credit: ESA/Astrium). Right: AWG chip manufactured for innoFSPEC.
}
   \end{figure}

The motivation was driven by the consideration that mass and volume are major cost drivers for space missions, hence miniaturizing an H-band spectrograph would be a worthwhile objective. Fig.~\ref{fig:1} illustrates the potential savings by showing the prominent example of the NIRspec spectrograph on board of the James Webb Space Telescope (JWST). The instrument has a size of 1900~mm~$\times$~1400~mm~$\times$~700~mm, and a mass of 196~kg. For reasons of thermal and mechanical stability, the mirror mounts and the optical bench base plate are manufactured out of silicon carbide ceramic and already amount to a mass of 100~kg, i.e. half of the total mass of the instrument. It is then quite obvious that shrinking appreciable parts of the free space optical system to a PIC would immediately result in reduced size, mass, and potentially improved thermal and mechanical stability.

\clearpage

\subsection{First experiments and exploration}
\label{subsec:exploration}

The very first steps to design and manufacture AWGs for spectroscopy were conducted in collaboration with the Leibniz Institute for High Performance Microelectronics (IHP) [\citenum{Fernando2012a,Fernando2012b}]. Initially, a conventional AWG, however without output receiver waveguides, was designed on a silicon nitride/SiO2/Si (Si3N4-SiO2-Si) platform for its relatively high refractive index, which, for a given channel spacing, was expected to allow a more compact device than Silica-on-Silicon. CMOS compatibility and the presence of a high non-linear optical coefficient was considered to be an added advantage for the manufacture of a densely integrated photonic chip, that could potentially also include a ring resonator based frequency comb for on-chip wavelength calibration. An analytical calculation, based on Gaussian beam approximation, was used to determine the optimal flat plane position where the non-uniformity in 1/e electric field widths would be minimal, to become the diced output plane for further coupling to a cross-dispersion optics and an imaging camera. The AWG was designed
to resolve 48 spectral channels with 0.4~nm (50GHz) resolution and adjacent channel cross-talk level within a 0.2nm window (ITU-grid) ~ -28dB. The calculated insertion loss non-uniformity was close to 3dB. The footprint of the high index contrast ($\Delta$n=23\%) area was 12$\times$8.5 mm$^2$. The modelled mean spectral resolving power at the flat image-plane was computed to R = 7600. 

   \begin{figure} [h]
   \begin{center}
    \begin{tabular}{c} 
   \includegraphics[height=45mm]{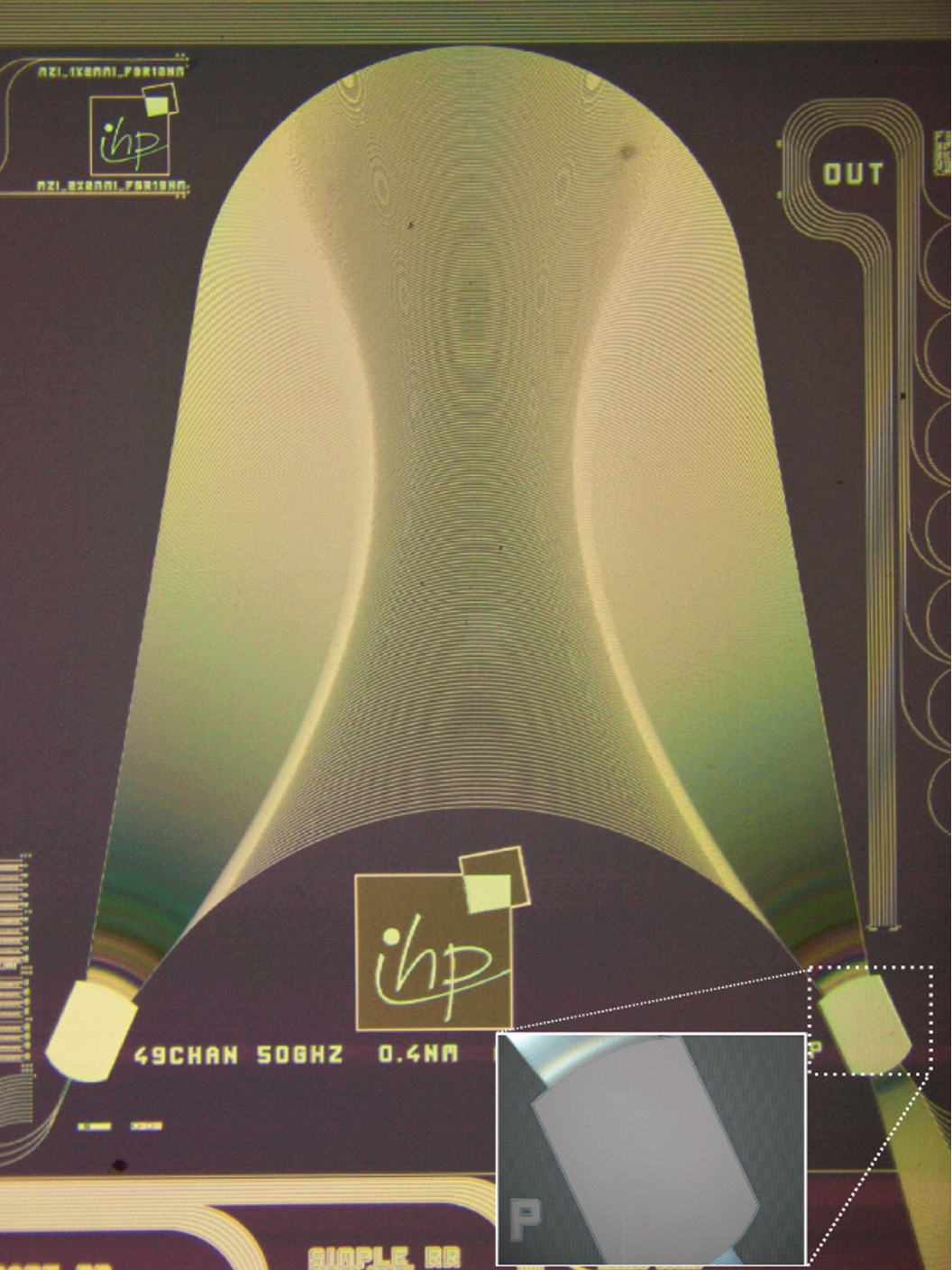}
	\end{tabular}
	\end{center}
   \caption[example] 
   { \label{fig:2} 
First experiments with AWGs on Si3N4 platform.
}
   \end{figure} 

The not entirely satisfactory experience with the silicon nitride platform led to a re-design using Silica-on-Silicon (SoS) technology, to tailor specific performance parameters of interest to a resolving power of R = $\lambda/\Delta\lambda$= 60,000 [\citenum{Fernando2014}].  The chip was designed to resolve up to 646 spectral lines per spectral order, with a wavelength spacing of 25 pm, at a central wavelength of 1630 nm. Fabricated test waveguides were stress engineered in order to compensate the inherent birefringence of SoS waveguides. The birefringence values of fabricated test structures were quantified, to be on the order of $10^{-6}$, the theoretical value required to avoid the formation of ghost-images, through inscription of Bragg-gratings on straight waveguides and subsequent measurement of Bragg-reflection spectra. An interferometer system was integrated on the same chip in order to allow for the characterization of phase errors of the waveguide array. Moreover, promising results of first fabricated key photonics components to form other complex integrated photonic circuits, such as astro-interferometers, using silicon nitride-on-insulator (SNOI) technology were also presented. The fabricated PICs included multimode interference based devices (power splitter/combiners, optical cross/bar-switches), directional-couplers with varying power ratios, Mach-Zehnder interferometers, and the AWG. First results of annealed, low-hydrogen SNOI based devices were promising and comparable to SOI and commercial devices, with device excess-loss less than 2 dB and under 1 dB/cm waveguide-loss at NIR-wavelengths. 

Prompted by the latter step, the design of a folded-architecture AWG for the astronomical H-band with a theoretical maximum resolving power R = 60,000 at 1630 nm was investigated [\citenum{Stoll2017}]. The geometry of the device was optimized for a compact structure with a footprint of 55~mm~$\times$~39.3~mm on SiO2 platform. To evaluate the fabrication challenges of such high-resolution AWGs, the effects of random perturbations of the effective refractive index (RI) distribution in the free propagation region (FPR), and small variations of the array waveguide optical lengths were numerically investigated. The results of the study have shown a dramatic degradation of the point spread function (PSF) for a random effective RI distribution with variance values above $10^{-4}$ for both the FPR and the waveguide array. Based on these results, requirements on the fabrication technology for high-resolution AWG-based spectrographs were derived.

Furthermore, a numerical and experimental study of the impact of phase errors on the performance of large, high-resolution arrayed waveguide gratings (AWG) for applications in astronomy was conducted [\citenum{Stoll2020a}]. Using a scalar diffraction model, the transmission spectrum of an AWG under random variations of the optical waveguide lengths was studied. Phase error correction was simulated by numerically trimming the lengths of the optical waveguides to the nearest integer multiple of the central wavelength. The optical length error distribution of a custom-fabricated silica AWG was measured using frequency-domain interferometry and Monte-Carlo fitting of interferogram intensities. As a conclusion, the estimate for the phase-error limited size of a waveguide array manufactured using state-of-the-art technology was presented. It was also shown that post-processing can, in principle, eliminate phase errors as a performance limiting factor of AWGs for astronomical spectroscopy.

\subsection{Fabricated AWGs Generation~I}
\label{subsecGenI}
Based on the previous studies, the fabrication of a first generation of custom-developed AWG on a silica platform for spectroscopic applications in near-infrared astronomy was attempted [\citenum{Stoll2021}]. This work was targeting a comprehensive description of the design, numerical simulation and characterization of several different types of devices, aiming at spectral resolving powers of 15,000–60,000 in the astronomical H-band. The AWGs were fabricated in a foundry run by Enablence Technologies Inc. using an atmospheric pressure chemical vapor deposition process (APCVD). The spectral characteristics of the fabricated devices were investigaged in terms of insertion loss and estimated spectral resolving power and compared with the  numerical simulations of the design. The different AWG types had been designed with resolving powers of up to 18,900 from the output channel 3-dB transmission bandwidth. Based on a first characterization results, two of the best performing AWGs were selected for further processing by removal of the output waveguide array and polishing the output facet to optical quality. The diced output is needed for integration as the primary diffractive element in a cross-dispersed spectrograph. For the selected devices, the imaging properties were measured with regards to spectral resolution in direct imaging mode, geometry-related defocus aberration, and polarization sensitivity of the spectral image. This work identified phase error control, birefringence control, and aberration suppression as the major issues to be solved towards high performance integrated photonics spectrographs.

   \begin{figure} [h]
   \begin{center}
    \begin{tabular}{c} 
   \includegraphics[height=8cm]{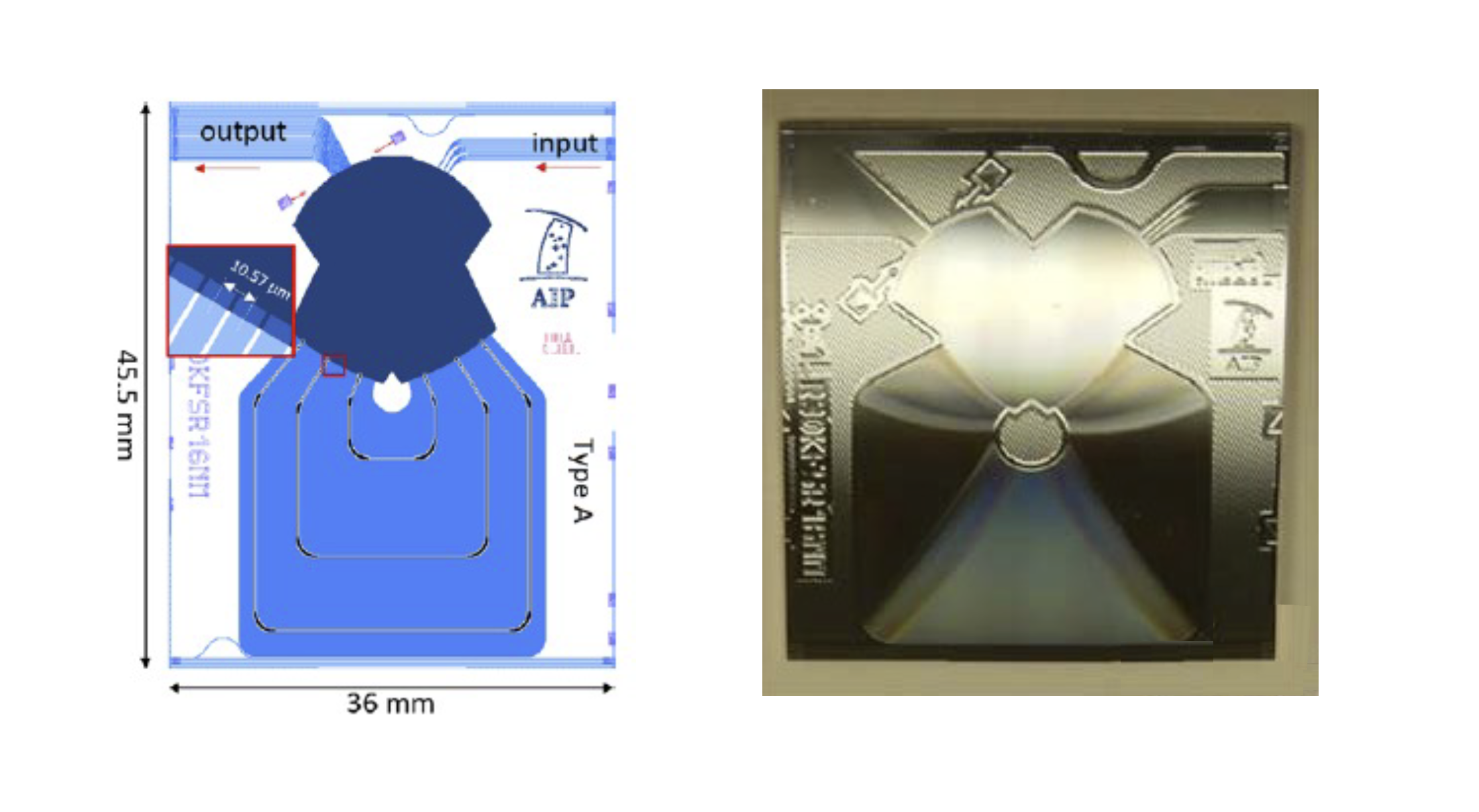}
	\end{tabular}
	\end{center}
   \caption[example] 
   { \label{fig:3} 
AWG Gen I fabricated on SoS platform.
}
   \end{figure} 

\subsection{Fabricated AWGs Generation~II}
\label{subsecGenII}

With lessons learned from the design, manufacture, and characterization of Gen I AWGs, an improved variant was developed using the three-stigmatic-point method with the goal to solve the problem of degraded image quality along the flat output facet of the devices as a result of longitudinal aberration [\citenum{Stoll2021a}]. This work presented theoretical and experimental results on the design, simulation and characterization of optimized Gen II silica-on-silicon AWGs.
Several mid-to-high resolution field-flattened AWG designs were derived, targeting resolving powers of 11,000 - 35,000 in the astronomical H-band, by iterative computation of differential coefficients of the optical path function. Numerical simulations were used to study the imaging properties of the designs in a wide wavelength range between 1500 nm and 1680 nm. The design-specific degradation of spectral resolving power at far-off-centre wavelengths was discussed, and possible solutions suggested. Fig.~\ref{fig:3} shows the lithographic mask that was used to manufacture a variety of different sizes and types of AWGs, based on the optimized design. The photographs show three designs of different geometry. The wafers were again fabricated by Enablence using APCVD like for Gen I. Seven selected devices were characterized in the lab, with performance parameters listed in the table shown in Fig.~\ref{fig:3}. The results show that it is possible to achieve resolving powers of around R = 30.000, and insertion losses as low as 1.45 dB.

   \begin{figure} [h]
   \begin{center}
    \begin{tabular}{c} 
   \includegraphics[height=9cm]{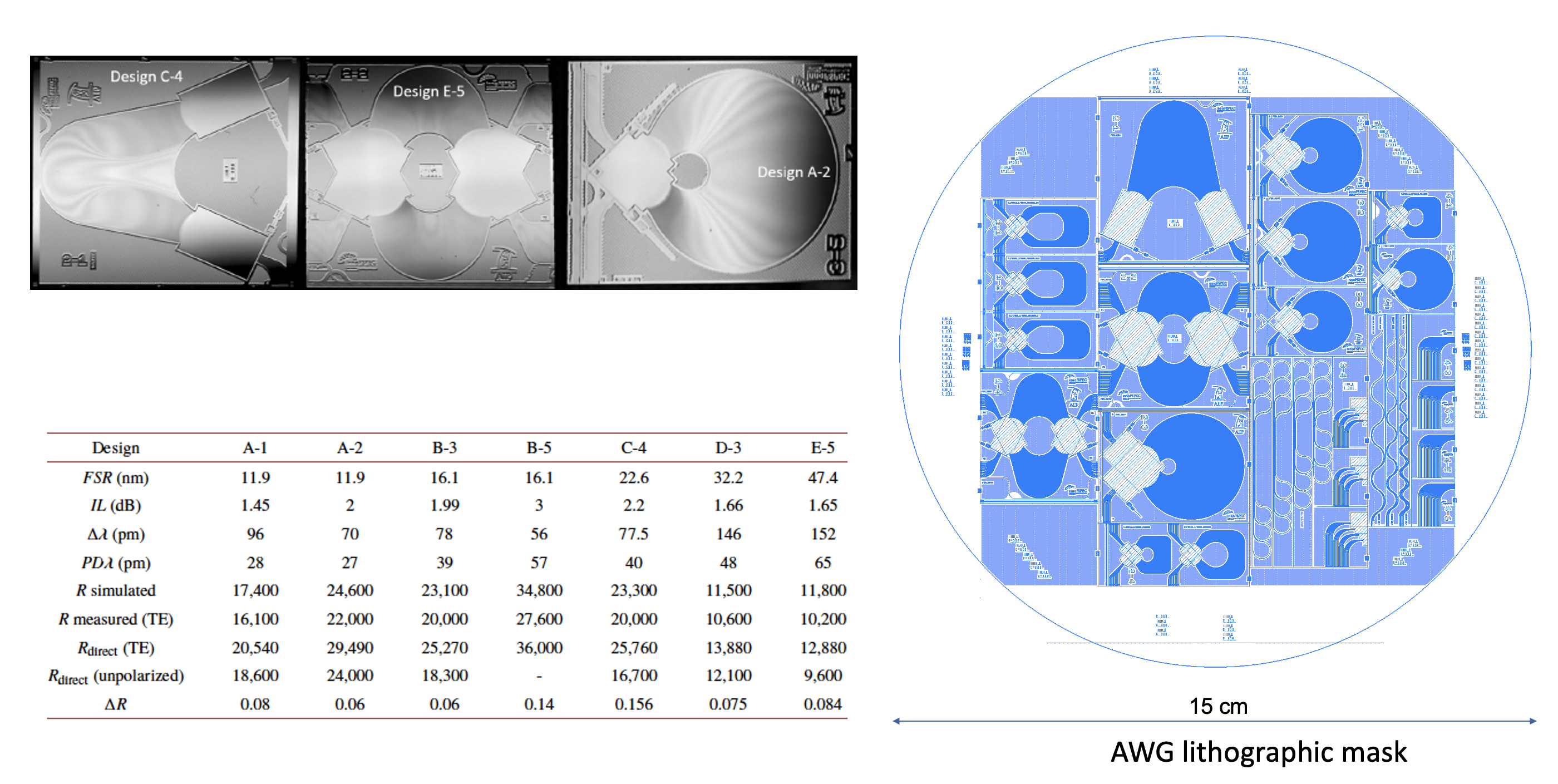}
	\end{tabular}
	\end{center}
   \caption[example] 
   { \label{fig:4} 
AWG Gen II fabricated on SoS platform. Right: lithographic mask for the 6 inch wafer containing different types of AWG, fabricated by Enablence. Upper left: photograph of three types of AWG with different geometries. Lower left: table with performance parameters as obtained from lab characterization.
}
   \end{figure} 

Beyond the initial exploratory work and Gen I and II designs, we mention for completeness without detailed discussion further work on echelle gratings [\citenum{Stoll2020}], and the peculiar design for an AWG that is embedded as a helical structure within an optical fiber, rather than on a planar chip [\citenum{Stoll2019}].

\clearpage

\subsection{PAWS, the Potsdam Arrayed Waveguide Spectrograph}
\label{subsecPAWS}

\begin{figure} [h]
   \begin{center}
    \begin{tabular}{c} 
   \includegraphics[height=60mm]{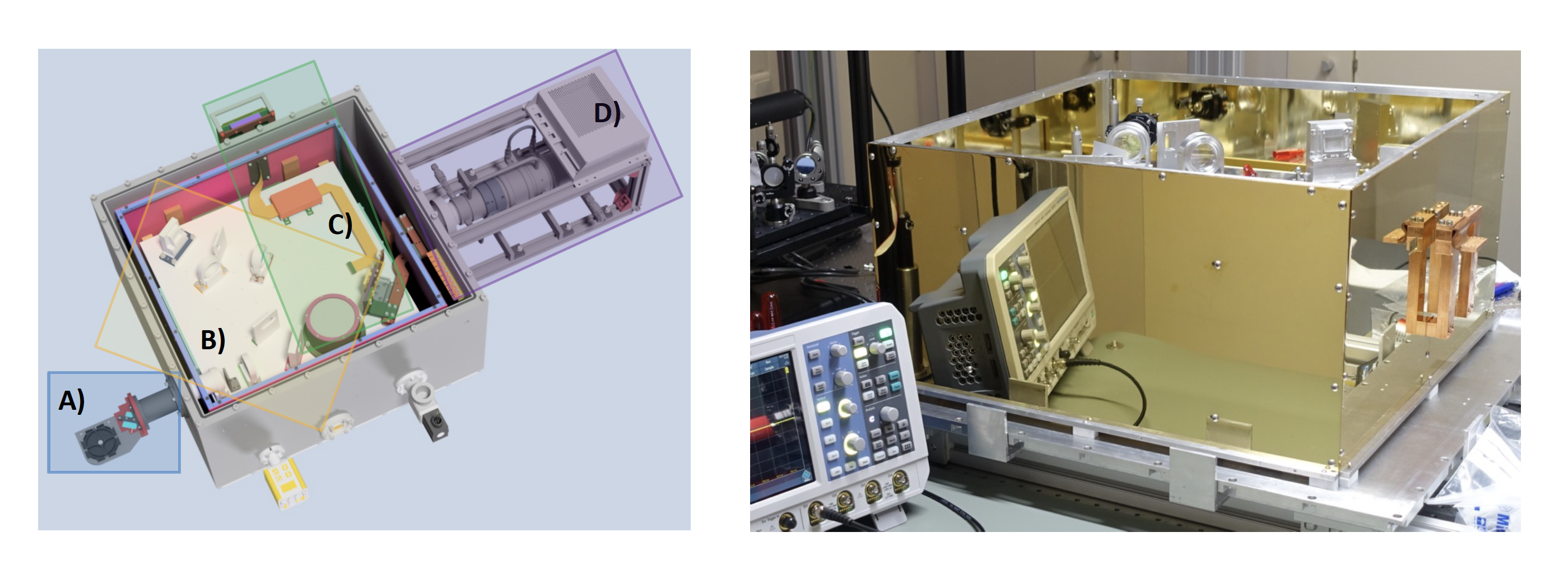}
	\end{tabular}
	\end{center}
   \caption[example] 
   { \label{fig:5} 
AWG demonstrator instrument PAWS. Left: CAD layout, showing AWG mount A with microscopy objective, connecting to the cryostat (rectangular box). Inside of the cryostat, the free space optical system for cross-dispersion B is cooled down to a temperature of 140 K. The detector C, cooled to 80 K, is an MCT NIR array H2RG from Teledyne. Cooling is accomplished through the cryocooler D. Right: open cryostat, showing gold-plated radiation shield that reduces radiative thermal losses to a minimum.
}
   \end{figure}

A crucial step to enable progress for any new technology in astronomical instrumenation is the need to validate the performance with an on-sky experiment that can be used to demonstrate technology readiness levels TRL~6 and TRL~7. To this end, innoFSPEC has designed and built an instrument, the Potsdam Arrayed Waveguide Spectrograph (PAWS) [\citenum{Hernandez2020,Hernandez2022}]. Fig.~\ref{fig:5} illustrates the major components, including the Gen II AWG chip as described in Section~\ref{subsecGenII}, that is mounted in subsystem A. Commissioning at the telescope and on-sky testing is planned at Calar Alto Observatory in southern Spain.

\section{PIC ring resonators in astronomical frequency combs for precision wavelength calibration}
\label{sec:rrcomb}

Optical frequency combs have been instrumental for high precision and accuracy in wavelength calibration for astronomical high resolution spectrographs, most notably achieving Doppler velocity measurements of stars that are accurate to about 1~m/s. Such accuracies are necessary for the spectroscopic detection of exo-planets. However, even spectrographs with medium or even low resolution would benefit from properties of frequency combs that deliver equidistant emission lines of comparable intensity - properties that are not at all warranted by classical spectral line lamps. However, the cost of commercially available laser frequency combs is prohibitively high. innoFSPEC has made attempts to find solutions that would address affordable general purpose frequency combs, however not necessarily at high spectral resolution, or the highest possible accuracy. 

Early on, two different platforms (and approaches) were numerically and experimentally investigated, targeting medium and low resolution spectrographs at astronomical facilities [\citenum{Chavez2012}]. 

In the first approach, a frequency comb was generated by propagating two lasers through three nonlinear stages. The first two stages serve for the generation of low-noise ultra-short pulses, while the final stage is a low-dispersion highly-nonlinear fibre where the pulses undergo strong spectral broadening. The wavelength of one of the lasers can be tuned, allowing the comb line spacing being continuously varied during the calibration procedure. The input power, the dispersion, the nonlinear coefficient, and fibre lengths in the nonlinear stages were defined and optimized by solving the generalized nonlinear Schr\"odinger equation. Experimentally,  the 250~GHz line-spacing frequency comb was generated by using two narrow linewidth lasers that were adiabatically compressed, firstly, in a standard fibre and, secondly, in a double-clad Er/Yb doped fibre. The spectral broadening finally took place in a highly nonlinear fibre resulting in an astro-comb with 250 calibration lines (covering a bandwidth of 500 nm) with good spectral equalization. 

The second approach aimed to generate optical frequency combs in dispersion-optimized silicon nitride ring resonators. A technique for lowering and flattening the chromatic dispersion in silicon nitride waveguides with silica cladding was proposed and demonstrated. By minimizing the waveguide dispersion over a broader wavelength range two goals were targeted: enhancing the phase matching for non-linear interactions, and broadening the wavelength coverage for the generated frequency combs.

For this purpose, instead of one cladding layer, the design incorporated two layers with appropriate thicknesses. It was possible to demonstrate nearly zero dispersion (with +/- 4 ps/nm-km variation) over the spectral region from 1.4 to 2.3 $\mu$m. 

The first approach was demonstrated with an experiment using a fiber-fed copy of a spectrograph module from the Multi Unit Spectroscopic Explorer (MUSE) [\citenum{Bacon2010}]. The frequency comb was generated by propagating two free-running lasers at 1554.3 and 1558.9 nm through two dispersion optimized nonlinear fibers [\citenum{Chavez2014}]. The generated comb was centered at 1590 nm and comprised more than one hundred lines with an optical-signal-to-noise ratio larger than 30 dB. A nonlinear crystal was used to frequency double the whole comb spectrum, which was efficiently converted into the 800 nm spectral band. A series of tests demonstrated not only that the comb delivered significantly more emission lines than a comparison Neon spectral line lamp, but also that equidistancy of the comb lines was confirmed with an absolute accuracy of 0.4 pm. 

Furthermore, on-sky tests were conducted using the Potsdam Multi-Aperture Spectrograph (PMAS) [\citenum{Roth2005}] at the 3.5 m Calar Alto Telescope [\citenum{Chavez2018}].

   \begin{figure} [h]
   \begin{center}
    \begin{tabular}{c} 
   \includegraphics[height=55mm]{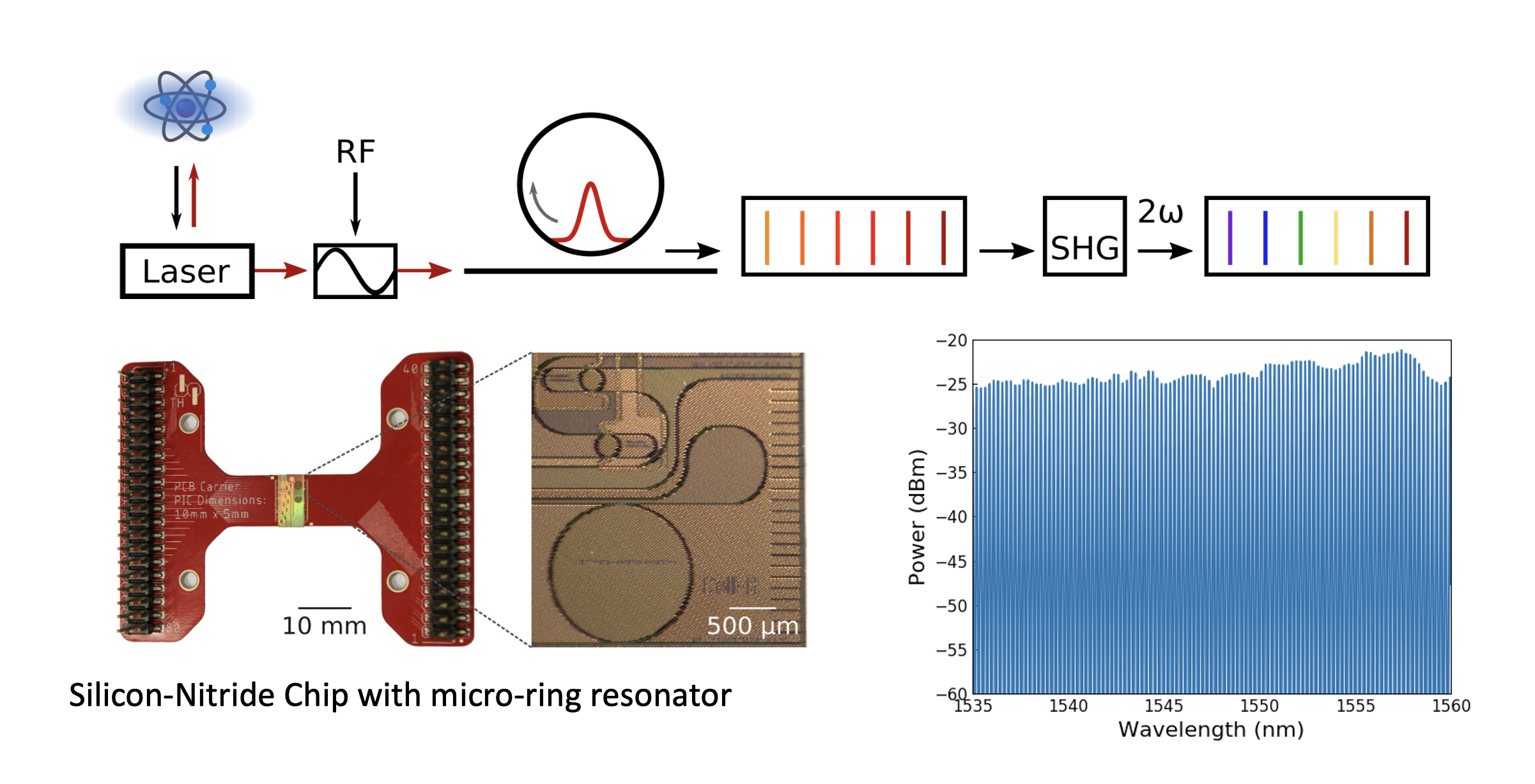}
	\end{tabular}
	\end{center}
   \caption[example] 
   { \label{fig:6} 
Frequency comb based on micro ring resonator.
}
   \end{figure} 

The second approach was subsequently studied, and first results reported for frequency combs in a silicon nitride micro-ring resonator, achieving an ultra-stable repetition frequency of 28.55 GHz [\citenum{Bodenmueller2020}], Fig.~\ref{fig:6}. The combs were generated by means of an amplitude modulated pump laser at 1568.8 nm and compared to numerical calculation based on a modified Lugiato-Lefever-Equation. The comb spectrum at a power level of -40 dB with respect to the pump line was found to span a wavelength range of 70~nm in this first experiment. 

   \begin{figure} [h]
   \begin{center}
    \begin{tabular}{c} 
   \includegraphics[height=40mm]{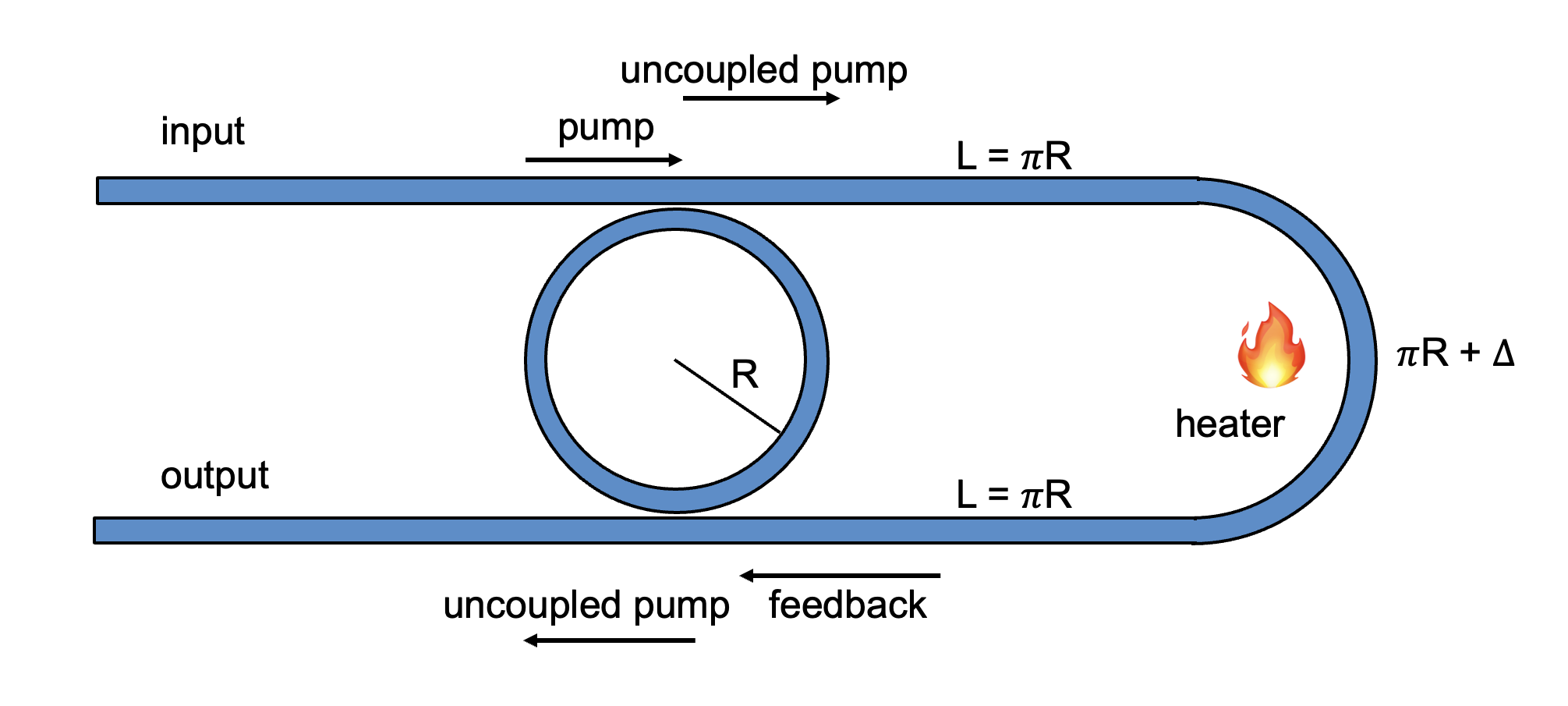}
	\end{tabular}
	\end{center}
   \caption[example] 
   { \label{fig:7} 
Micro-ring resonator with interferometric feedback loop, controlled by a micro heater, after [\citenum{Chavez2022}].
}
   \end{figure} 

Further research resulted in a break-through towards practical applications, in particular a viable route towards developing a turn-key system for reliable operation at an observatory [\citenum{Chavez2022}]. Nonlinear Kerr micro-resonators have enabled the understanding of dissipative solitons and their application to optical frequency comb generation. However, the conversion efficiency of the pump power into a soliton frequency comb typically remains below a few percent. A solution for the problem of low efficiency consists in the implementation of a hybrid Mach-Zehnder ring resonator geometry, that is realized as a micro-ring resonator embedded in an additional cavity with twice the optical path length of the ring, see Fig.~\ref{fig:7}. The resulting interferometric back coupling enables an unprecedented control of the pump depletion: pump-to-frequency comb conversion efficiencies of up to 55\% of the input pump power have been experimentally demonstrated. The robustness of the novel on-chip geometry was verified through manufacture and testing a large variety of dissipative Kerr soliton combs. Micro-resonators with feedback enable new regimes of coherent soliton comb generation with applications in astronomy, spectroscopy and telecommunications. The concept is presently being implemented in a prototype instrument, the Potsdam Frequency Comb (POCO).

   \begin{figure} [h]
   \begin{center}
    \begin{tabular}{c} 
   \includegraphics[height=70mm]{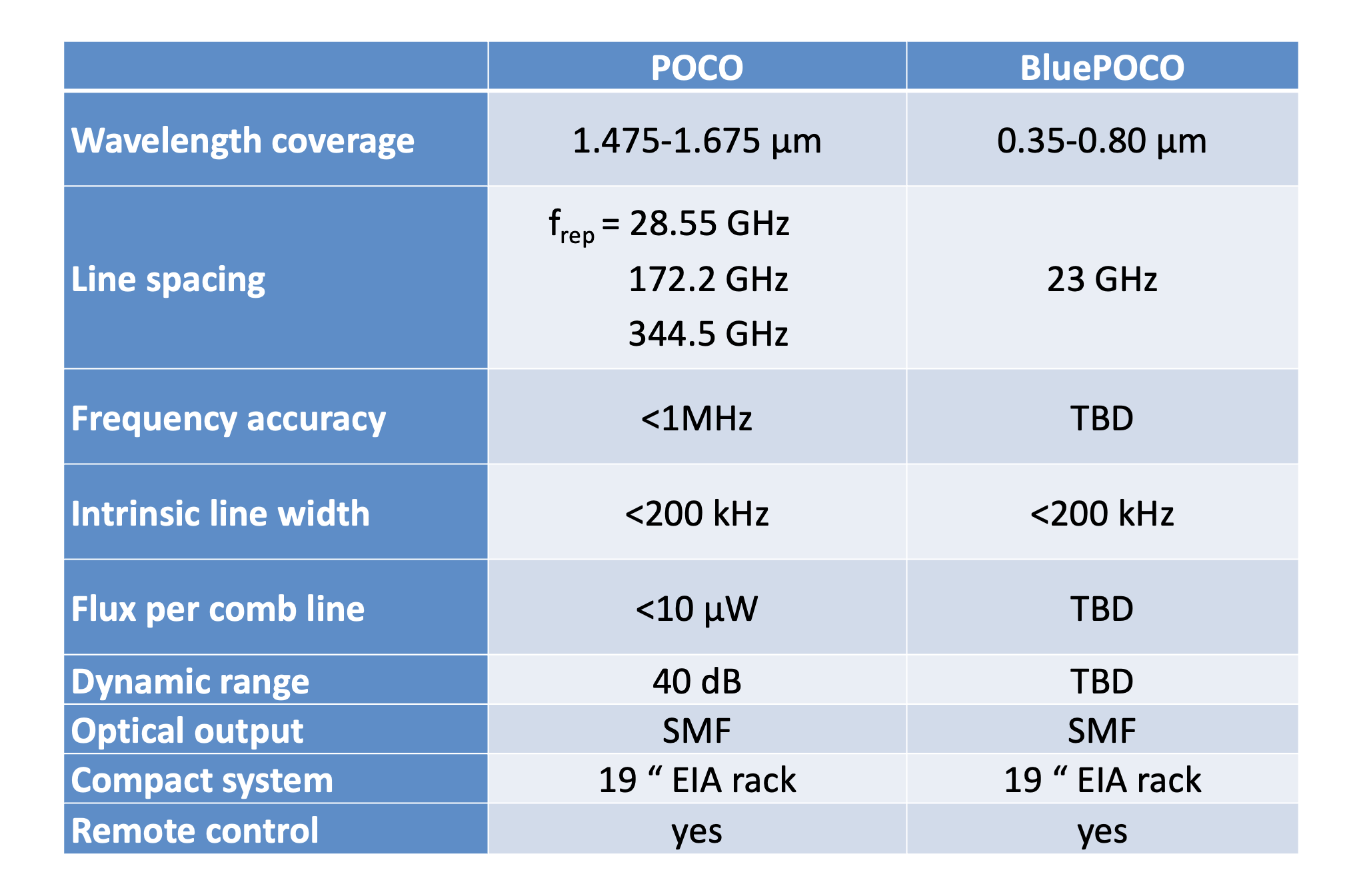}
	\end{tabular}
	\end{center}
   \caption[example] 
   { \label{fig:8} 
Parameters of the Potsdam Frequency Comb POCO, and a future version for use at optical/blue wavelengths.
}
   \end{figure} 

\section{Discrete beam combiners (DBC) for large astronomical interferometers}
\label{sec:DBC}

\begin{figure} [h]
   \begin{center}
    \begin{tabular}{c} 
   \includegraphics[height=80mm]{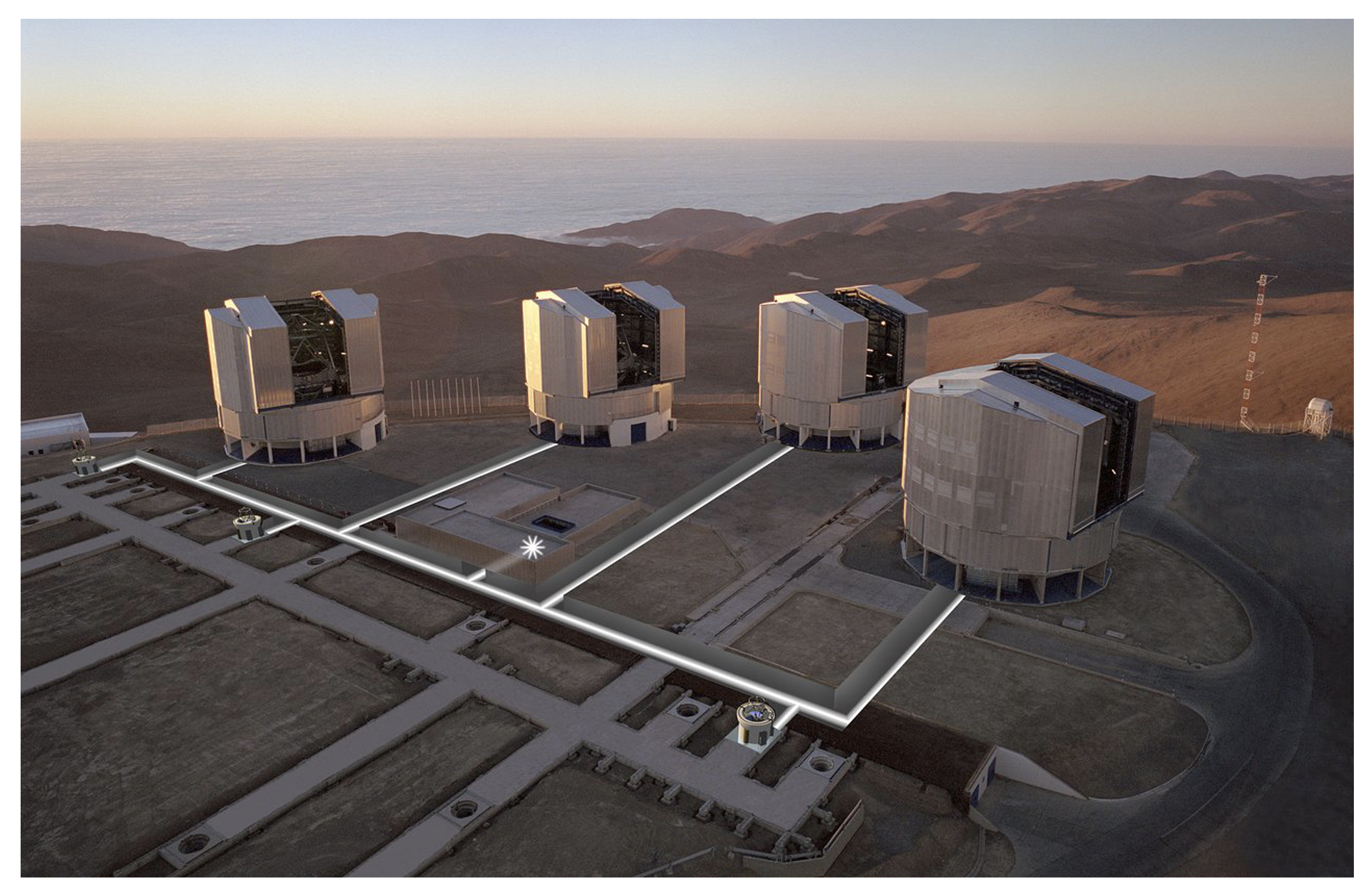}
	\end{tabular}
	\end{center}
   \caption[example] 
   { \label{fig:9} 
Very Large Telescope Interferometer (VLTI). Credit: ESO.
}
   \end{figure}

\begin{figure} [h]
   \begin{center}
    \begin{tabular}{c} 
   \includegraphics[height=70mm]{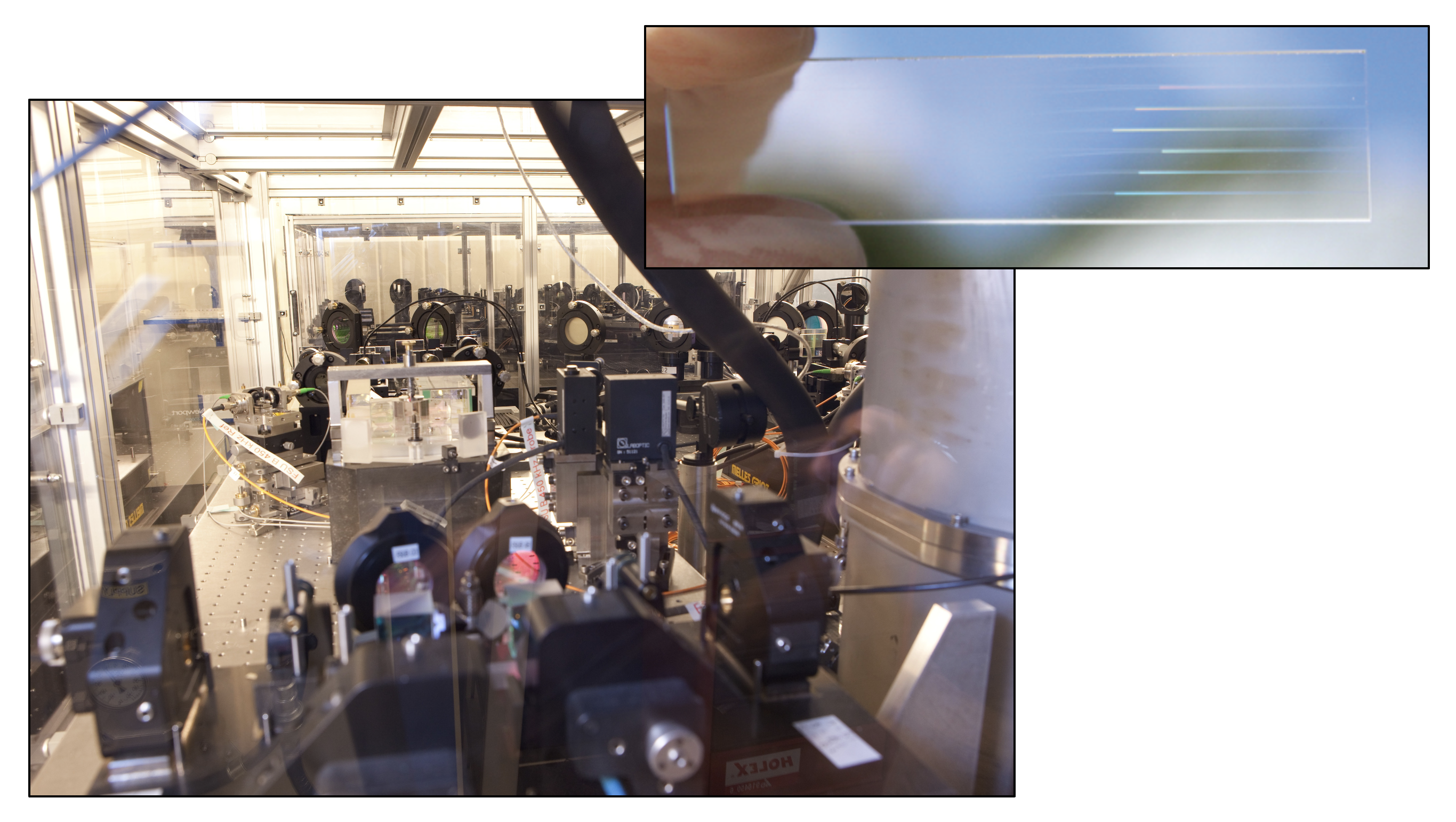}
	\end{tabular}
	\end{center}
   \caption[example] 
   { \label{fig:10} 
VLTI free space optics beam combiner. Credit: ESO. Insert: DBC glass plate for comparison, fabricated for innoFSPEC, and tested at WHT.
}
   \end{figure}

Arguably, the first PICs ever appearing for astronomical instrumentation, were used in NIR interferometers, such as the European Southern Observatory's Very Large Telescope Interferometer (VLTI) in Chile. An aerial view of the VLTI platform is shown in Fig.~\ref{fig:9}, where the channels and delay lines that connect the four Unit Telescopes and auxiliary telescopes are highlighted in white color. The location where the various beams are combined to create the interferometric fringes is indicated with the symbol of a white star. The beam combiner laboratory in its classical configuration harbors a plethora of free space optical elements like mirrors and lenses that obviously must be controlled thermally and mechanically to maintain their positions during the course of an observation to a fraction of a wavelength. If the optical path length varies in the different trains of the setup, the visibility of the fringes will be compromised.  Fig.~\ref{fig:10} is an illustration of this situation, suggesting that a tiny integrated optics beam combiner would solve a multitude of issues, concerning complexity, alignment, and stability. After initial demonstration of feasibility [\citenum{Malbet1999}], integrated optics beam combiners have advanced to 2nd generation VLTI instrumentation [\citenum{Benisty2009}], most prominently for the instrument GRAVITY [\citenum{Abuter2017}]. Amongst other targets, GRAVITY has been used to observe the massive black hole in the center of the Milky Way. Perhaps one of the most striking results is the measurement of gravitational redshift of the star S2 in the orbit around the Galactic center black hole [\citenum{Abuter2018}], but there are many other measurements that make the instrument an outstanding facility. There is no doubt that the GRAVITY integrated optics beam combiner has become an essential element of the success story of this instrument. However, the technology of the device is limited in so far as it is fabricated as a planar chip whose geometry constrains the liberty of routing waveguides for four or more input beams. innoFSPEC has engaged in developing Discrete Beam Combiners (DBC) that depart from the planar geometry and allow for three-dimensional routing of waveguides in glass plates thanks to inscription with femto-second lasers, also known as Ultra-fast Laser Inscription (ULI) [\citenum{Minardi2010,Minardi2016}]. An example of a DBC is shown in the upper right of Fig.~\ref{fig:10}.

\begin{figure} [h]
   \begin{center}
    \begin{tabular}{c} 
   \includegraphics[height=70mm]{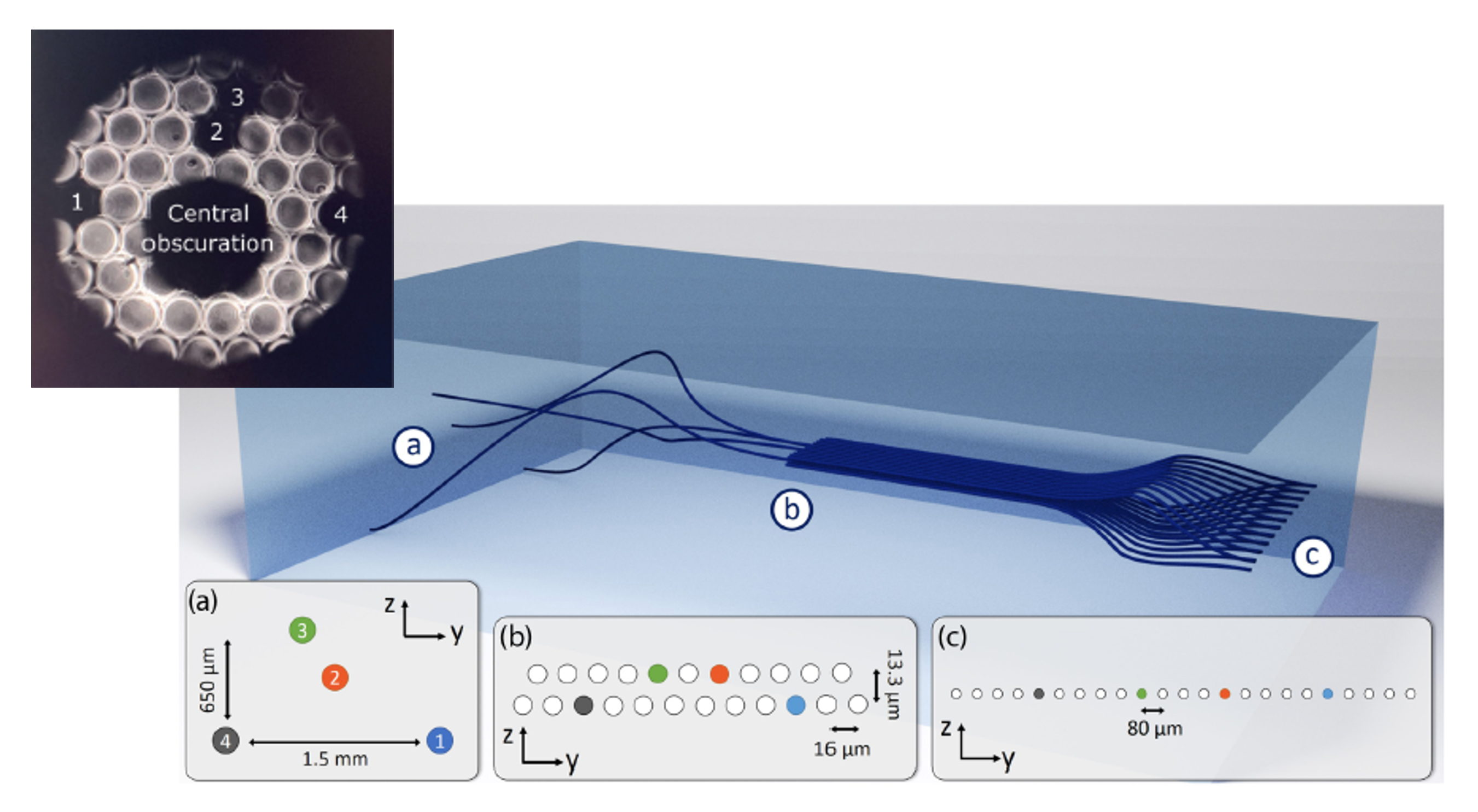}
	\end{tabular}
	\end{center}
   \caption[example] 
   { \label{fig:11} 
Four-beam discrete beam combiner for on-sky test at William Herschel Telescope, La Palma.
}
   \end{figure}

A four-beam DBC has been developed [\citenum{Nayak2020}] for the astronomical H-band and tested on-sky at the 4.2m William Herschel Telescope, La Palma (WHT) [\citenum{Nayak2021}]. As opposed to the example discussed above for GRAVITY, where light is combined that comes from different telescopes, the experiment consists of a four-input pupil remapper, followed by a DBC and a 23-output reformatter. It was devised to test aperture masking -- a technique that has originally been developed for single 8-10m class telescopes with the goal of diffraction limited imaging, delivering superior angular resolution in comparison to speckle interferometry, or adaptive optics. The DBC has been manufactured by Politecnico di Milano in an alumino-borosilicate substrate using ULI. A cavity-dumped Yb:KYW laser operating at 1030~nm with 300 fs pulses and a repetition rate of 1~MHz was employed for the process. In the experiment at WHT, the device was operated with a deformable mirror
and a microlens array, whose purpose was to select sub-pupils of the telescope pupil, as shown in the upper left of Fig.~\ref{fig:11}, labelled 1, 2, 3, 4. The corresponding beams were injected into the DBC in plane (a) as highlighted by the colored symbols 1,$\ldots$,4. The corresponding waveguides were routed in three dimensions to accurately maintain the designed optical path lenghts before entering the beam combiner section proper, that is arranged in a zig-zag pattern of 23 evanescently coupling waveguides as shown in section (b). Four of these waveguides are actively fed from the input section, whereas nearest and second nearest neigboring waveguides accomplish the interference, which is read out in the reformatted linear array as shown for section (c). The intensity pattern on the output is then sensing the phases, to be recorded conveniently with a direct imaging camera. The on-sky experiment 
has measured visibility amplitudes and closure phases obtained on the bright stars Vega and Altair. While the coherence function could be indeed reconstructed, the results showed significant dispersion from
the expected values due to the only limited signal-to-noise ratio that could be obtained, considering the limited light collecting area at a 4m class telescope. Nonetheless, the experiment has been an important first validation step towards future application for long-baseline interferometry. While this work was conducted in the H-band, further work with collaborators in the UK, Germany, and in the U.S.A. have led to similar on-sky tests with DBCs in the K-band, using the CHARA interferometer on Mt. Wilson Observatory [\citenum{Harris2020, Pike2020,Siliprandi2022}].

\section{Aperiodic fiber Bragg gratings for complex astronomical filters}
\label{sec:FBG} 

While conclusive results are as yet not available, we mention for completeness first activities at innoFSPEC to experiment with ULI inscription in photonic chips for the purpose of creating aperiodic fiber Bragg gratings (FBG) that have been demonstrated previously to enable OH suppression on the basis of single-mode fibers 
[\citenum{Bland-Hawthorn2011}],[\citenum{BlandHawthorn2016,Trinh2012,Trinh2013,Ellis2018,Ellis2020,Ellis2020a}]. The goal of OH suppression is to filter out extremely bright atmospheric OH emission lines that limit severely the ability of large ground-based telescopes to record NIR spectra of faint galaxies.  While FBG have been reported to be feasible with ULI inscription in optical fibers [\citenum{Goebel2018}], it would be interesting to demonstrate the same in photonic chips. A competing technology is currently under development using phase mask inscription with ultraviolet laser light
[\citenum{Rahman2020}].

\section{Conclusions}
\label{sec:conclusions}
Photonic integrated circuits are presenting interesting capabilities for instrumentation in astronomy. They are already established key components for long-baseline interferometry in the NIR. We have demonstrated custom-made PICs for integrated photonic spectrographs and frequency combs, now proceeding to prototypes, hence TRL6. DBC have been developed and demonstrated for the astronomical H-band, with progress being made towards longer wavelengths in the K-band. Silica-on-Silicon, silcon nitride, and ULI inscribed glass substrates are the technology platforms that have been employed successfully. Astrophotonics can be considered a firmly established sub-discipline with interesting prospects of further growth and development.

\acknowledgments       
We acknowledge support from BMBF grants 03Z22AN11 ''Astrophotonics", 03Z22A511 ''astrOOptics'', 03Z22AB1A ''NIR-DETECT'', and 03Z22AI1 "Strategic Investment" through the BMBF program ''Unternehmen Region'', and DFG grant 326946494, "NAIR", at the Zentrum für Innovationskompetenz innoFSPEC.
\bibliography{report} 

\begin{thebibliography}{10}

\bibitem{Galilei1610}
{Galilei}, G.,  [{\em {Sidereus nuncius}}{\nolinebreak\hspace{0.1em}]} (1610).

\bibitem{Angel1977}
{Angel}, J.~R.~P., {Adams}, M.~T., {Boroson}, T.~A., and {Moore}, R.~L., ``{A
  very large optical telescope array linked with fused silica fibers.},'' {\em
  ApJ}~{\bf 218},  776--782 (Dec. 1977).

\bibitem{Angel1979}
{Hubbard}, E.~N., {Angel}, J.~R.~P., and {Gresham}, M.~S., ``{Operation of a
  long fused silica fiber as a link between telescope and spectrograph.},''
  {\em ApJ}~{\bf 229},  1074--1078 (May 1979).

\bibitem{Hill1980}
{Hill}, J.~M., {Angel}, J.~R.~P., {Scott}, J.~S., {Lindley}, D., and {Hintzen},
  P., ``{Multiple object spectroscopy: the medusa spectrograph.},'' {\em
  ApJL}~{\bf 242},  L69--L72 (Dec. 1980).

\bibitem{Gunn2006}
{Gunn}, J.~E., {Siegmund}, W.~A., {Mannery}, E.~J., {Owen}, R.~E., {Hull},
  C.~L., {Leger}, R.~F., {Carey}, L.~N., {Knapp}, G.~R., {York}, D.~G.,
  {Boroski}, W.~N., {Kent}, S.~M., {Lupton}, R.~H., {Rockosi}, C.~M., {Evans},
  M.~L., {Waddell}, P., {Anderson}, J.~E., {Annis}, J., {Barentine}, J.~C.,
  {Bartoszek}, L.~M., {Bastian}, S., {Bracker}, S.~B., {Brewington}, H.~J.,
  {Briegel}, C.~I., {Brinkmann}, J., {Brown}, Y.~J., {Carr}, M.~A.,
  {Czarapata}, P.~C., {Drennan}, C.~C., {Dombeck}, T., {Federwitz}, G.~R.,
  {Gillespie}, B.~A., {Gonzales}, C., {Hansen}, S.~U., {Harvanek}, M., {Hayes},
  J., {Jordan}, W., {Kinney}, E., {Klaene}, M., {Kleinman}, S.~J., {Kron},
  R.~G., {Kresinski}, J., {Lee}, G., {Limmongkol}, S., {Lindenmeyer}, C.~W.,
  {Long}, D.~C., {Loomis}, C.~L., {McGehee}, P.~M., {Mantsch}, P.~M.,
  {Neilsen}, Eric~H., J., {Neswold}, R.~M., {Newman}, P.~R., {Nitta}, A.,
  {Peoples}, John, J., {Pier}, J.~R., {Prieto}, P.~S., {Prosapio}, A.,
  {Rivetta}, C., {Schneider}, D.~P., {Snedden}, S., and {Wang}, S.-i., ``{The
  2.5 m Telescope of the Sloan Digital Sky Survey},'' {\em AJ}~{\bf 131},
  2332--2359 (Apr. 2006).

\bibitem{Mayor2003}
{Mayor}, M., {Pepe}, F., {Queloz}, D., {Bouchy}, F., {Rupprecht}, G., {Lo
  Curto}, G., {Avila}, G., {Benz}, W., {Bertaux}, J.~L., {Bonfils}, X., {Dall},
  T., {Dekker}, H., {Delabre}, B., {Eckert}, W., {Fleury}, M., {Gilliotte}, A.,
  {Gojak}, D., {Guzman}, J.~C., {Kohler}, D., {Lizon}, J.~L., {Longinotti}, A.,
  {Lovis}, C., {Megevand}, D., {Pasquini}, L., {Reyes}, J., {Sivan}, J.~P.,
  {Sosnowska}, D., {Soto}, R., {Udry}, S., {van Kesteren}, A., {Weber}, L., and
  {Weilenmann}, U., ``{Setting New Standards with HARPS},'' {\em The
  Messenger}~{\bf 114},  20--24 (Dec. 2003).

\bibitem{Strassmeier2015}
{Strassmeier}, K.~G., {Ilyin}, I., {J{\"a}rvinen}, A., {Weber}, M., {Woche},
  M., {Barnes}, S.~I., {Bauer}, S.~M., {Beckert}, E., {Bittner}, W.,
  {Bredthauer}, R., {Carroll}, T.~A., {Denker}, C., {Dionies}, F., {DiVarano},
  I., {D{\"o}scher}, D., {Fechner}, T., {Feuerstein}, D., {Granzer}, T.,
  {Hahn}, T., {Harnisch}, G., {Hofmann}, A., {Lesser}, M., {Paschke}, J.,
  {Pankratow}, S., {Plank}, V., {Pl{\"u}schke}, D., {Popow}, E., and
  {Sablowski}, D., ``{PEPSI: The high-resolution {\'e}chelle spectrograph and
  polarimeter for the Large Binocular Telescope},'' {\em Astronomische
  Nachrichten}~{\bf 336},  324 (May 2015).

\bibitem{Pepe2021}
{Pepe}, F., {Cristiani}, S., {Rebolo}, R., {Santos}, N.~C., {Dekker}, H.,
  {Cabral}, A., {Di Marcantonio}, P., {Figueira}, P., {Lo Curto}, G., {Lovis},
  C., {Mayor}, M., {M{\'e}gevand}, D., {Molaro}, P., {Riva}, M., {Zapatero
  Osorio}, M.~R., {Amate}, M., {Manescau}, A., {Pasquini}, L., {Zerbi}, F.~M.,
  {Adibekyan}, V., {Abreu}, M., {Affolter}, M., {Alibert}, Y., {Aliverti}, M.,
  {Allart}, R., {Allende Prieto}, C., {{\'A}lvarez}, D., {Alves}, D., {Avila},
  G., {Baldini}, V., {Bandy}, T., {Barros}, S.~C.~C., {Benz}, W., {Bianco}, A.,
  {Borsa}, F., {Bourrier}, V., {Bouchy}, F., {Broeg}, C., {Calderone}, G.,
  {Cirami}, R., {Coelho}, J., {Conconi}, P., {Coretti}, I., {Cumani}, C.,
  {Cupani}, G., {D'Odorico}, V., {Damasso}, M., {Deiries}, S., {Delabre}, B.,
  {Demangeon}, O.~D.~S., {Dumusque}, X., {Ehrenreich}, D., {Faria}, J.~P.,
  {Fragoso}, A., {Genolet}, L., {Genoni}, M., {G{\'e}nova Santos}, R.,
  {Gonz{\'a}lez Hern{\'a}ndez}, J.~I., {Hughes}, I., {Iwert}, O., {Kerber}, F.,
  {Knudstrup}, J., {Landoni}, M., {Lavie}, B., {Lillo-Box}, J., {Lizon}, J.~L.,
  {Maire}, C., {Martins}, C.~J.~A.~P., {Mehner}, A., {Micela}, G.,
  {Modigliani}, A., {Monteiro}, M.~A., {Monteiro}, M.~J.~P.~F.~G., {Moschetti},
  M., {Murphy}, M.~T., {Nunes}, N., {Oggioni}, L., {Oliveira}, A., {Oshagh},
  M., {Pall{\'e}}, E., {Pariani}, G., {Poretti}, E., {Rasilla}, J.~L.,
  {Rebord{\~a}o}, J., {Redaelli}, E.~M., {Santana Tschudi}, S., {Santin}, P.,
  {Santos}, P., {S{\'e}gransan}, D., {Schmidt}, T.~M., {Segovia}, A.,
  {Sosnowska}, D., {Sozzetti}, A., {Sousa}, S.~G., {Span{\`o}}, P., {Su{\'a}rez
  Mascare{\~n}o}, A., {Tabernero}, H., {Tenegi}, F., {Udry}, S., and {Zanutta},
  A., ``{ESPRESSO at VLT. On-sky performance and first results},'' {\em
  A\&A}~{\bf 645},  A96 (Jan. 2021).

\bibitem{Roth2005}
{Roth}, M.~M., {Kelz}, A., {Fechner}, T., {Hahn}, T., {Bauer}, S.-M., {Becker},
  T., {B{\"o}hm}, P., {Christensen}, L., {Dionies}, F., {Paschke}, J., {Popow},
  E., {Wolter}, D., {Schmoll}, J., {Laux}, U., and {Altmann}, W., ``{PMAS: The
  Potsdam Multi-Aperture Spectrophotometer. I. Design, Manufacture, and
  Performance},'' {\em PASP}~{\bf 117},  620--642 (June 2005).

\bibitem{Kelz2006}
{Kelz}, A., {Verheijen}, M. A.~W., {Roth}, M.~M., {Bauer}, S.~M., {Becker}, T.,
  {Paschke}, J., {Popow}, E., {S{\'a}nchez}, S.~F., and {Laux}, U., ``{PMAS:
  The Potsdam Multi-Aperture Spectrophotometer. II. The Wide Integral Field
  Unit PPak},'' {\em PASP}~{\bf 118},  129--145 (Jan. 2006).

\bibitem{Sanchez2012}
{S{\'a}nchez}, S.~F., {Kennicutt}, R.~C., {Gil de Paz}, A., {van de Ven}, G.,
  {V{\'\i}lchez}, J.~M., {Wisotzki}, L., {Walcher}, C.~J., {Mast}, D.,
  {Aguerri}, J.~A.~L., {Albiol-P{\'e}rez}, S., {Alonso-Herrero}, A., {Alves},
  J., {Bakos}, J., {Bart{\'a}kov{\'a}}, T., {Bland-Hawthorn}, J., {Boselli},
  A., {Bomans}, D.~J., {Castillo-Morales}, A., {Cortijo-Ferrero}, C., {de
  Lorenzo-C{\'a}ceres}, A., {Del Olmo}, A., {Dettmar}, R.~J., {D{\'\i}az}, A.,
  {Ellis}, S., {Falc{\'o}n-Barroso}, J., {Flores}, H., {Gallazzi}, A.,
  {Garc{\'\i}a-Lorenzo}, B., {Gonz{\'a}lez Delgado}, R., {Gruel}, N., {Haines},
  T., {Hao}, C., {Husemann}, B., {Igl{\'e}sias-P{\'a}ramo}, J., {Jahnke}, K.,
  {Johnson}, B., {Jungwiert}, B., {Kalinova}, V., {Kehrig}, C., {Kupko}, D.,
  {L{\'o}pez-S{\'a}nchez}, {\'A}.~R., {Lyubenova}, M., {Marino}, R.~A.,
  {M{\'a}rmol-Queralt{\'o}}, E., {M{\'a}rquez}, I., {Masegosa}, J., {Meidt},
  S., {Mendez-Abreu}, J., {Monreal-Ibero}, A., {Montijo}, C., {Mour{\~a}o},
  A.~M., {Palacios-Navarro}, G., {Papaderos}, P., {Pasquali}, A., {Peletier},
  R., {P{\'e}rez}, E., {P{\'e}rez}, I., {Quirrenbach}, A., {Rela{\~n}o}, M.,
  {Rosales-Ortega}, F.~F., {Roth}, M.~M., {Ruiz-Lara}, T.,
  {S{\'a}nchez-Bl{\'a}zquez}, P., {Sengupta}, C., {Singh}, R., {Stanishev}, V.,
  {Trager}, S.~C., {Vazdekis}, A., {Viironen}, K., {Wild}, V., {Zibetti}, S.,
  and {Ziegler}, B., ``{CALIFA, the Calar Alto Legacy Integral Field Area
  survey. I. Survey presentation},'' {\em A\&A}~{\bf 538},  A8 (Feb. 2012).

\bibitem{Bundy2015}
{Bundy}, K., {Bershady}, M.~A., {Law}, D.~R., {Yan}, R., {Drory}, N.,
  {MacDonald}, N., {Wake}, D.~A., {Cherinka}, B., {S{\'a}nchez-Gallego}, J.~R.,
  {Weijmans}, A.-M., {Thomas}, D., {Tremonti}, C., {Masters}, K., {Coccato},
  L., {Diamond-Stanic}, A.~M., {Arag{\'o}n-Salamanca}, A., {Avila-Reese}, V.,
  {Badenes}, C., {Falc{\'o}n-Barroso}, J., {Belfiore}, F., {Bizyaev}, D.,
  {Blanc}, G.~A., {Bland-Hawthorn}, J., {Blanton}, M.~R., {Brownstein}, J.~R.,
  {Byler}, N., {Cappellari}, M., {Conroy}, C., {Dutton}, A.~A., {Emsellem}, E.,
  {Etherington}, J., {Frinchaboy}, P.~M., {Fu}, H., {Gunn}, J.~E., {Harding},
  P., {Johnston}, E.~J., {Kauffmann}, G., {Kinemuchi}, K., {Klaene}, M.~A.,
  {Knapen}, J.~H., {Leauthaud}, A., {Li}, C., {Lin}, L., {Maiolino}, R.,
  {Malanushenko}, V., {Malanushenko}, E., {Mao}, S., {Maraston}, C.,
  {McDermid}, R.~M., {Merrifield}, M.~R., {Nichol}, R.~C., {Oravetz}, D.,
  {Pan}, K., {Parejko}, J.~K., {Sanchez}, S.~F., {Schlegel}, D., {Simmons}, A.,
  {Steele}, O., {Steinmetz}, M., {Thanjavur}, K., {Thompson}, B.~A., {Tinker},
  J.~L., {van den Bosch}, R. C.~E., {Westfall}, K.~B., {Wilkinson}, D.,
  {Wright}, S., {Xiao}, T., and {Zhang}, K., ``{Overview of the SDSS-IV MaNGA
  Survey: Mapping nearby Galaxies at Apache Point Observatory},'' {\em
  ApJ}~{\bf 798},  7 (Jan. 2015).

\bibitem{Bland-Hawthorn2004}
{Bland-Hawthorn}, J., ``{Astrophotonics comes of age: an OH-suppressing
  infrared fibre},'' {\em Anglo-Australian Observatory Epping Newsletter}~{\bf
  106},  4 (Dec. 2004).

\bibitem{Norris2019}
{Norris}, B. and {Bland-Hawthorn}, J., ``{Astrophotonics: The Rise of
  Integrated Photonics in Astronomy},'' {\em Optics \& Photonics News}~{\bf
  30},  26 (May 2019).

\bibitem{Minardi2021}
{Minardi}, S., {Harris}, R.~J., and {Labadie}, L., ``{Astrophotonics: astronomy
  and modern optics},'' {\em AAPR}~{\bf 29},  6 (Dec. 2021).

\bibitem{Roth2008}
{Roth}, M.~M., {L{\"o}hmannsr{\"o}ben}, H.-G., {Kelz}, A., and {Kumke}, M.,
  ``{innoFSPEC: fiber optical spectroscopy and sensing},'' in [{\em Advanced
  Optical and Mechanical Technologies in Telescopes and
  Instrumentation}{\nolinebreak\hspace{0.1em}]},  {Atad-Ettedgui}, E. and
  {Lemke}, D., eds., {\em Society of Photo-Optical Instrumentation Engineers
  (SPIE) Conference Series} {\bf 7018},  70184X (July 2008).

\bibitem{Haynes2010}
{Haynes}, R., {Reich}, O., {Rambold}, W., {Hass}, R., and {Janssen}, K.,
  ``{Fibre optical spectroscopy and sensing innovation at innoFSPEC Potsdam},''
  in [{\em Modern Technologies in Space- and Ground-based Telescopes and
  Instrumentation}{\nolinebreak\hspace{0.1em}]},  {Atad-Ettedgui}, E. and
  {Lemke}, D., eds., {\em Society of Photo-Optical Instrumentation Engineers
  (SPIE) Conference Series} {\bf 7739},  77394J (July 2010).

\bibitem{Gatkine2019}
{Gatkine}, P., {Veilleux}, S., {Mather}, J., {Betters}, C., {Bland-Hawthorn},
  J., {Bryant}, J., {Cenko}, S.~B., {Dagenais}, M., {Deming}, D., {Ellis}, S.,
  {Greenhouse}, M., {Harris}, A., {Jovanovic}, N., {Kuhlmann}, S., {Kutyrev},
  A., {Leon-Saval}, S., {Madhav}, K., {Moseley}, S., {Norris}, B., {Rauscher},
  B., {Roth}, M., and {Vogel}, S., ``{State of the Profession:
  Astrophotonics},'' in [{\em Bulletin of the American Astronomical
  Society}{\nolinebreak\hspace{0.1em}]},   {\bf 51},  285 (Sept. 2019).

\bibitem{Cheben2007}
{Cheben}, P., {Schmid}, J.~H., {Del{\^a}ge}, A., {Densmore}, A., {Janz}, S.,
  {Lamontagne}, B., {Lapointe}, J., {Post}, E., {Waldron}, P., and {Xu}, D.~X.,
  ``{A high-resolution silicon-on-insulator arrayed waveguide grating
  microspectrometer with sub-micrometer aperture waveguides},'' {\em Optics
  Express}~{\bf 15},  2299--2306 (Mar. 2007).

\bibitem{Bland-Hawthorn2006}
{Bland-Hawthorn}, J. and {Horton}, A., ``{Instruments without optics: an
  integrated photonic spectrograph},'' in [{\em Ground-based and Airborne
  Instrumentation for Astronomy}{\nolinebreak\hspace{0.1em}]},  {McLean}, I.~S.
  and {Iye}, M., eds., {\em Society of Photo-Optical Instrumentation Engineers
  (SPIE) Conference Series} {\bf 6269},  62690N (June 2006).

\bibitem{Cvetojevic2009}
{Cvetojevic}, N., {Lawrence}, J.~S., {Ellis}, S.~C., {Bland-Hawthorn}, J.,
  {Haynes}, R., and {Horton}, A., ``{Characterization and on-sky demonstration
  of an integrated photonic spectrograph for astronomy},'' {\em Optics
  Express}~{\bf 17},  18643--18650 (Oct. 2009).

\bibitem{Cvetojevic2012}
{Cvetojevic}, N., {Jovanovic}, N., {Betters}, C., {Lawrence}, J.~S., {Ellis},
  S.~C., {Robertson}, G., and {Bland-Hawthorn}, J., ``{First starlight spectrum
  captured using an integrated photonic micro-spectrograph},'' {\em A\&A}~{\bf
  544},  L1 (Aug. 2012).

\bibitem{Bland-Hawthorn2010}
{Bland-Hawthorn}, J., {Lawrence}, J., {Robertson}, G., {Campbell}, S., {Pope},
  B., {Betters}, C., {Leon-Saval}, S., {Birks}, T., {Haynes}, R., {Cvetojevic},
  N., and {Jovanovic}, N., ``{PIMMS: photonic integrated multimode
  microspectrograph},'' in [{\em Ground-based and Airborne Instrumentation for
  Astronomy III}{\nolinebreak\hspace{0.1em}]},  {McLean}, I.~S., {Ramsay},
  S.~K., and {Takami}, H., eds., {\em Society of Photo-Optical Instrumentation
  Engineers (SPIE) Conference Series} {\bf 7735},  77350N (July 2010).

\bibitem{Fernando2012a}
{Fernando}, H. N.~J., {Stoll}, A., {Boggio}, J.~C., {Haynes}, R., and {Roth},
  M.~M., ``{Arrayed waveguide gratings beyond communication: utilization of
  entire image-plane of output star-coupler for spectroscopy and sensing},'' in
  [{\em Silicon Photonics and Photonic Integrated Circuits
  III}{\nolinebreak\hspace{0.1em}]},  {Vivien}, L., {Honkanen}, S.~K.,
  {Pavesi}, L., and {Pelli}, S., eds., {\em Society of Photo-Optical
  Instrumentation Engineers (SPIE) Conference Series} {\bf 8431},  84311U (June
  2012).

\bibitem{Fernando2012b}
{Fernando}, H. N.~J., {Stoll}, A., {Eisermann}, R., {Boggio}, J.~C.,
  {Zimmermann}, L., {Haynes}, R., and {Roth}, M.~M., ``{Planar integrated
  photonics spectrograph on silicon-nitride-on-insulator: densely integrated
  systems for astrophotonics and spectroscopy},'' in [{\em Modern Technologies
  in Space- and Ground-based Telescopes and Instrumentation
  II}{\nolinebreak\hspace{0.1em}]},  {Navarro}, R., {Cunningham}, C.~R., and
  {Prieto}, E., eds., {\em Society of Photo-Optical Instrumentation Engineers
  (SPIE) Conference Series} {\bf 8450},  845046 (Sept. 2012).

\bibitem{Fernando2014}
{Fernando}, H. N.~J., {Stoll}, A., {Cvetojevic}, N., {Eisemann}, R.,
  {Tharanga}, N., {Holmes}, C., {B{\"o}hm}, M., {Roth}, M.~M., {Haynes}, R.,
  and {Zimmermann}, L., ``{Interferometers and spectrographs on
  silicon-platform for astrophysics: trends of astrophotonics},'' in [{\em
  Advances in Optical and Mechanical Technologies for Telescopes and
  Instrumentation}{\nolinebreak\hspace{0.1em}]},  {Navarro}, R., {Cunningham},
  C.~R., and {Barto}, A.~A., eds., {\em Society of Photo-Optical
  Instrumentation Engineers (SPIE) Conference Series} {\bf 9151},  915148 (July
  2014).

\bibitem{Stoll2017}
{Stoll}, A., {Zhang}, Z., {Haynes}, R., and {Roth}, M., ``{High-Resolution
  Arrayed-Waveguide-Gratings in Astronomy: Design and Fabrication
  Challenges},'' {\em Photonics for Solar Energy Systems IX}~{\bf 4},  30 (Apr.
  2017).

\bibitem{Stoll2020a}
{Stoll}, A., {Madhav}, K., and {Roth}, M., ``{Performance limits of
  astronomical arrayed waveguide gratings on a silica platform},'' {\em Optics
  Express}~{\bf 28},  39354 (Dec. 2020).

\bibitem{Stoll2021}
{Stoll}, A., {Madhav}, K.~V., and {Roth}, M.~M., ``{Design, simulation and
  characterization of integrated photonic spectrographs for astronomy:
  generation-I AWG devices based on canonical layouts},'' {\em Optics
  Express}~{\bf 29},  24947 (Aug. 2021).

\bibitem{Stoll2021a}
{Stoll}, A., {Madhav}, K., and {Roth}, M., ``{Design, simulation and
  characterization of integrated photonic spectrographs for astronomy II:
  low-aberration Generation-II AWG devices with three stigmatic points},'' {\em
  Optics Express}~{\bf 29},  36226 (Oct. 2021).

\bibitem{Stoll2020}
{Stoll}, A., {Wang}, Y., {Madhav}, K., and {Roth}, M., ``{Integrated
  ech{\'e}lle gratings for astrophotonics},'' in [{\em Advances in Optical
  Astronomical Instrumentation 2019}{\nolinebreak\hspace{0.1em}]},  {Ellis},
  S.~C. and {d'Orgeville}, C., eds., {\em Society of Photo-Optical
  Instrumentation Engineers (SPIE) Conference Series} {\bf 11203},  112030Z
  (Jan. 2020).

\bibitem{Stoll2019}
{Stoll}, A., {Zhang}, Z., {Sun}, K., {Madhav}, K., {Fiebrandt}, J., and {Roth},
  M.~M., ``{Cross-dispersed in-fibre spectrometer based on helix core
  bundle},'' {\em Journal of Modern Optics}~{\bf 66},  829--834 (May 2019).

\bibitem{Hernandez2020}
{Hernandez}, E., {Stoll}, A., {Bauer}, S.-M., {Berdja}, A., {Bernardi}, R.,
  {Guzman}, D., {G{\"u}nther}, A., {Madhav}, K., {Roth}, M.~M., {Sandin}, C.,
  and {Villanueva}, C., ``{Optomechanical design of PAWS, the Potsdam Arrayed
  Waveguide Spectrograph},'' in [{\em Society of Photo-Optical Instrumentation
  Engineers (SPIE) Conference Series}{\nolinebreak\hspace{0.1em}]},  {\em
  Society of Photo-Optical Instrumentation Engineers (SPIE) Conference Series}
  {\bf 11451},  114515O (Dec. 2020).

\bibitem{Hernandez2022}
{Hernandez}, E., {G{\"u}nther}, A., {Bauer}, S.-M., {Guzm{\'a}n}, C.~D.,
  {Sandin}, C., {Stoll}, A., {Vjesnica}, S., {Madhav}, K., and {Roth}, M.~M.,
  ``{System integration of the Potsdam Arrayed Waveguide Spectrograph
  (PAWS)},'' in [{\em Ground-based and Airborne Instrumentation for Astronomy
  IX}{\nolinebreak\hspace{0.1em}]},  {Evans}, C.~J., {Bryant}, J.~J., and
  {Motohara}, K., eds., {\em Society of Photo-Optical Instrumentation Engineers
  (SPIE) Conference Series} {\bf 12184},  121841O (Aug. 2022).

\bibitem{Chavez2012}
{Chavez Boggio}, J.~M., {Fremberg}, T., {Bodenm{\"u}ller}, D., {Wysmolek}, M.,
  {Sanyic}, H., {Fernando}, H., {Neumann}, J., {Kracht}, D., {Haynes}, R., and
  {Roth}, M.~M., ``{Astronomical optical frequency comb generation in nonlinear
  fibres and ring resonators: optimization studies},'' in [{\em Modern
  Technologies in Space- and Ground-based Telescopes and Instrumentation
  II}{\nolinebreak\hspace{0.1em}]},  {Navarro}, R., {Cunningham}, C.~R., and
  {Prieto}, E., eds., {\em Society of Photo-Optical Instrumentation Engineers
  (SPIE) Conference Series} {\bf 8450},  84501H (Sept. 2012).

\bibitem{Bacon2010}
{Bacon}, R., {Accardo}, M., {Adjali}, L., {Anwand}, H., {Bauer}, S., {Biswas},
  I., {Blaizot}, J., {Boudon}, D., {Brau-Nogue}, S., {Brinchmann}, J.,
  {Caillier}, P., {Capoani}, L., {Carollo}, C.~M., {Contini}, T., {Couderc},
  P., {Daguis{\'e}}, E., {Deiries}, S., {Delabre}, B., {Dreizler}, S.,
  {Dubois}, J., {Dupieux}, M., {Dupuy}, C., {Emsellem}, E., {Fechner}, T.,
  {Fleischmann}, A., {Fran{\c{c}}ois}, M., {Gallou}, G., {Gharsa}, T.,
  {Glindemann}, A., {Gojak}, D., {Guiderdoni}, B., {Hansali}, G., {Hahn}, T.,
  {Jarno}, A., {Kelz}, A., {Koehler}, C., {Kosmalski}, J., {Laurent}, F., {Le
  Floch}, M., {Lilly}, S.~J., {Lizon}, J.~L., {Loupias}, M., {Manescau}, A.,
  {Monstein}, C., {Nicklas}, H., {Olaya}, J.~C., {Pares}, L., {Pasquini}, L.,
  {P{\'e}contal-Rousset}, A., {Pell{\'o}}, R., {Petit}, C., {Popow}, E.,
  {Reiss}, R., {Remillieux}, A., {Renault}, E., {Roth}, M., {Rupprecht}, G.,
  {Serre}, D., {Schaye}, J., {Soucail}, G., {Steinmetz}, M., {Streicher}, O.,
  {Stuik}, R., {Valentin}, H., {Vernet}, J., {Weilbacher}, P., {Wisotzki}, L.,
  and {Yerle}, N., ``{The MUSE second-generation VLT instrument},'' in [{\em
  Ground-based and Airborne Instrumentation for Astronomy
  III}{\nolinebreak\hspace{0.1em}]},  {McLean}, I.~S., {Ramsay}, S.~K., and
  {Takami}, H., eds., {\em Society of Photo-Optical Instrumentation Engineers
  (SPIE) Conference Series} {\bf 7735},  773508 (July 2010).

\bibitem{Chavez2014}
{Chavez Boggio}, J.~M., {Fremberg}, T., {Moralejo}, B., {Rutowska}, M.,
  {Hernandez}, E., {Zajnulina}, M., {Kelz}, A., {Bodenm{\"u}ller}, D.,
  {Sandin}, C., {Wysmolek}, M., {Sayinc}, H., {Neumann}, J., {Haynes}, R., and
  {Roth}, M.~M., ``{Astronomical optical frequency comb generation and test in
  a fiber-fed MUSE spectrograph},'' in [{\em Advances in Optical and Mechanical
  Technologies for Telescopes and
  Instrumentation}{\nolinebreak\hspace{0.1em}]},  {Navarro}, R., {Cunningham},
  C.~R., and {Barto}, A.~A., eds., {\em Society of Photo-Optical
  Instrumentation Engineers (SPIE) Conference Series} {\bf 9151},  915120 (July
  2014).

\bibitem{Chavez2018}
{Chavez Boggio}, J.~M., {Fremberg}, T., {Bodenm{\"u}ller}, D., {Sandin}, C.,
  {Zajnulina}, M., {Kelz}, A., {Giannone}, D., {Rutowska}, M., {Moralejo}, B.,
  {Roth}, M.~M., {Wysmolek}, M., and {Sayinc}, H., ``{Wavelength calibration
  with PMAS at 3.5 m Calar Alto Telescope using a tunable astro-comb},'' {\em
  Optics Communications}~{\bf 415},  186--193 (May 2018).

\bibitem{Bodenmueller2020}
{Bodenm{\"u}ller}, D., {Chavez Boggio}, J.~M., and {Roth}, M.~M., ``{Optical
  frequency comb generated in micro-ring resonators by modulated pump-light},''
  in [{\em Advances in Optical Astronomical Instrumentation
  2019}{\nolinebreak\hspace{0.1em}]},  {Ellis}, S.~C. and {d'Orgeville}, C.,
  eds., {\em Society of Photo-Optical Instrumentation Engineers (SPIE)
  Conference Series} {\bf 11203},  112031H (Jan. 2020).

\bibitem{Chavez2022}
{Boggio}, J.~M.~C., {Bodenm{\"u}ller}, D., {Ahmed}, S., {Wabnitz}, S.,
  {Modotto}, D., and {Hansson}, T., ``{Efficient Kerr soliton comb generation
  in micro-resonator with interferometric back-coupling},'' {\em Nature
  Communications}~{\bf 13},  1292 (Mar. 2022).

\bibitem{Malbet1999}
{Malbet}, F., {Kern}, P., {Schanen-Duport}, I., {Berger}, J.~P.,
  {Rousselet-Perraut}, K., and {Benech}, P., ``{Integrated optics for
  astronomical interferometry. I. Concept and astronomical applications},''
  {\em A\&A Suppl}~{\bf 138},  135--145 (July 1999).

\bibitem{Benisty2009}
{Benisty}, M., {Berger}, J.~P., {Jocou}, L., {Labeye}, P., {Malbet}, F.,
  {Perraut}, K., and {Kern}, P., ``{An integrated optics beam combiner for the
  second generation VLTI instruments},'' {\em A\&A}~{\bf 498},  601--613 (May
  2009).

\bibitem{Abuter2017}
{GRAVITY Collaboration}, {Abuter}, R., {Accardo}, M., {Amorim}, A., {Anugu},
  N., {{\'A}vila}, G., {Azouaoui}, N., {Benisty}, M., {Berger}, J.~P., {Blind},
  N., {Bonnet}, H., {Bourget}, P., {Brandner}, W., {Brast}, R., {Buron}, A.,
  {Burtscher}, L., {Cassaing}, F., {Chapron}, F., {Choquet}, {\'E}.,
  {Cl{\'e}net}, Y., {Collin}, C., {Coud{\'e} Du Foresto}, V., {de Wit}, W., {de
  Zeeuw}, P.~T., {Deen}, C., {Delplancke-Str{\"o}bele}, F., {Dembet}, R.,
  {Derie}, F., {Dexter}, J., {Duvert}, G., {Ebert}, M., {Eckart}, A.,
  {Eisenhauer}, F., {Esselborn}, M., {F{\'e}dou}, P., {Finger}, G., {Garcia},
  P., {Garcia Dabo}, C.~E., {Garcia Lopez}, R., {Gendron}, E., {Genzel}, R.,
  {Gillessen}, S., {Gonte}, F., {Gordo}, P., {Grould}, M., {Gr{\"o}zinger}, U.,
  {Guieu}, S., {Haguenauer}, P., {Hans}, O., {Haubois}, X., {Haug}, M.,
  {Haussmann}, F., {Henning}, T., {Hippler}, S., {Horrobin}, M., {Huber}, A.,
  {Hubert}, Z., {Hubin}, N., {Hummel}, C.~A., {Jakob}, G., {Janssen}, A.,
  {Jochum}, L., {Jocou}, L., {Kaufer}, A., {Kellner}, S., {Kendrew}, S.,
  {Kern}, L., {Kervella}, P., {Kiekebusch}, M., {Klein}, R., {Kok}, Y., {Kolb},
  J., {Kulas}, M., {Lacour}, S., {Lapeyr{\`e}re}, V., {Lazareff}, B., {Le
  Bouquin}, J.~B., {L{\`e}na}, P., {Lenzen}, R., {L{\'e}v{\^e}que}, S.,
  {Lippa}, M., {Magnard}, Y., {Mehrgan}, L., {Mellein}, M., {M{\'e}rand}, A.,
  {Moreno-Ventas}, J., {Moulin}, T., {M{\"u}ller}, E., {M{\"u}ller}, F.,
  {Neumann}, U., {Oberti}, S., {Ott}, T., {Pallanca}, L., {Panduro}, J.,
  {Pasquini}, L., {Paumard}, T., {Percheron}, I., {Perraut}, K., {Perrin}, G.,
  {Pfl{\"u}ger}, A., {Pfuhl}, O., {Phan Duc}, T., {Plewa}, P.~M., {Popovic},
  D., {Rabien}, S., {Ram{\'\i}rez}, A., {Ramos}, J., {Rau}, C., {Riquelme}, M.,
  {Rohloff}, R.~R., {Rousset}, G., {Sanchez-Bermudez}, J., {Scheithauer}, S.,
  {Sch{\"o}ller}, M., {Schuhler}, N., {Spyromilio}, J., {Straubmeier}, C.,
  {Sturm}, E., {Suarez}, M., {Tristram}, K.~R.~W., {Ventura}, N., {Vincent},
  F., {Waisberg}, I., {Wank}, I., {Weber}, J., {Wieprecht}, E., {Wiest}, M.,
  {Wiezorrek}, E., {Wittkowski}, M., {Woillez}, J., {Wolff}, B., {Yazici}, S.,
  {Ziegler}, D., and {Zins}, G., ``{First light for GRAVITY: Phase referencing
  optical interferometry for the Very Large Telescope Interferometer},'' {\em
  A\&A}~{\bf 602},  A94 (June 2017).

\bibitem{Abuter2018}
{GRAVITY Collaboration}, {Abuter}, R., {Amorim}, A., {Anugu}, N.,
  {Baub{\"o}ck}, M., {Benisty}, M., {Berger}, J.~P., {Blind}, N., {Bonnet}, H.,
  {Brandner}, W., {Buron}, A., {Collin}, C., {Chapron}, F., {Cl{\'e}net}, Y.,
  {Coud{\'e} Du Foresto}, V., {de Zeeuw}, P.~T., {Deen}, C.,
  {Delplancke-Str{\"o}bele}, F., {Dembet}, R., {Dexter}, J., {Duvert}, G.,
  {Eckart}, A., {Eisenhauer}, F., {Finger}, G., {F{\"o}rster Schreiber}, N.~M.,
  {F{\'e}dou}, P., {Garcia}, P., {Garcia Lopez}, R., {Gao}, F., {Gendron}, E.,
  {Genzel}, R., {Gillessen}, S., {Gordo}, P., {Habibi}, M., {Haubois}, X.,
  {Haug}, M., {Hau{\ss}mann}, F., {Henning}, T., {Hippler}, S., {Horrobin}, M.,
  {Hubert}, Z., {Hubin}, N., {Jimenez Rosales}, A., {Jochum}, L., {Jocou}, K.,
  {Kaufer}, A., {Kellner}, S., {Kendrew}, S., {Kervella}, P., {Kok}, Y.,
  {Kulas}, M., {Lacour}, S., {Lapeyr{\`e}re}, V., {Lazareff}, B., {Le Bouquin},
  J.~B., {L{\'e}na}, P., {Lippa}, M., {Lenzen}, R., {M{\'e}rand}, A.,
  {M{\"u}ler}, E., {Neumann}, U., {Ott}, T., {Palanca}, L., {Paumard}, T.,
  {Pasquini}, L., {Perraut}, K., {Perrin}, G., {Pfuhl}, O., {Plewa}, P.~M.,
  {Rabien}, S., {Ram{\'\i}rez}, A., {Ramos}, J., {Rau}, C.,
  {Rodr{\'\i}guez-Coira}, G., {Rohloff}, R.~R., {Rousset}, G.,
  {Sanchez-Bermudez}, J., {Scheithauer}, S., {Sch{\"o}ller}, M., {Schuler}, N.,
  {Spyromilio}, J., {Straub}, O., {Straubmeier}, C., {Sturm}, E., {Tacconi},
  L.~J., {Tristram}, K.~R.~W., {Vincent}, F., {von Fellenberg}, S., {Wank}, I.,
  {Waisberg}, I., {Widmann}, F., {Wieprecht}, E., {Wiest}, M., {Wiezorrek}, E.,
  {Woillez}, J., {Yazici}, S., {Ziegler}, D., and {Zins}, G., ``{Detection of
  the gravitational redshift in the orbit of the star S2 near the Galactic
  centre massive black hole},'' {\em A\&A}~{\bf 615},  L15 (July 2018).

\bibitem{Minardi2010}
{Minardi}, S. and {Pertsch}, T., ``{Interferometric beam combination with
  discrete optics},'' {\em Optics Letters}~{\bf 35},  3009 (Aug. 2010).

\bibitem{Minardi2016}
{Minardi}, S., {Lacour}, S., {Berger}, J.-P., {Labadie}, L., {Thomson}, R.~R.,
  {Haniff}, C., and {Ireland}, M., ``{Beam combination schemes and technologies
  for the Planet Formation Imager},'' in [{\em Optical and Infrared
  Interferometry and Imaging V}{\nolinebreak\hspace{0.1em}]},  {Malbet}, F.,
  {Creech-Eakman}, M.~J., and {Tuthill}, P.~G., eds., {\em Society of
  Photo-Optical Instrumentation Engineers (SPIE) Conference Series} {\bf 9907},
   99071N (Aug. 2016).

\bibitem{Nayak2020}
{Nayak}, A.~S., {Piacentini}, S., {Sharma}, T.~K., {Corrielli}, G., {Osellame},
  R., {Labadie}, L., {Minardi}, S., {Pedretti}, E., {Madhav}, K., and {Roth},
  M.~M., ``{Integrated optics-interferometry using pupil remapping and beam
  combination at astronomical H-band},'' in [{\em Advances in Optical
  Astronomical Instrumentation 2019}{\nolinebreak\hspace{0.1em}]},  {Ellis},
  S.~C. and {d'Orgeville}, C., eds., {\em Society of Photo-Optical
  Instrumentation Engineers (SPIE) Conference Series} {\bf 11203},  112030V
  (Jan. 2020).

\bibitem{Nayak2021}
{Nayak}, A.~S., {Labadie}, L., {Sharma}, T.~K., {Piacentini}, S., {Corrielli},
  G., {Osellame}, R., {Gendron}, {\'E}., {Buey}, J.-T.~M., {Chemla}, F.,
  {Cohen}, M., {Bharmal}, N.~A., {Bardou}, L.~F., {Staykov}, L., {Osborn}, J.,
  {Morris}, T.~J., {Pedretti}, E., {Dinkelaker}, A.~N., {Madhav}, K.~V., and
  {Roth}, M.~M., ``{First stellar photons for an integrated optics discrete
  beam combiner at the William Herschel Telescope},'' {\em Appl. Opt.}~{\bf 60},  D129
  (July 2021).

\bibitem{Harris2020}
{Harris}, R.~J., {Sharma}, T.~K., {Davenport}, J.~J., {Hottinger}, P.,
  {Anagnos}, T., {Nayak}, A.~S., {Quirrenbach}, A., {Labadie}, L., {Madhav},
  K.~V., and {Roth}, M.~M., ``{NAIR: Novel Astronomical Instrumentation through
  photonic Reformatting},'' in [{\em Society of Photo-Optical Instrumentation
  Engineers (SPIE) Conference Series}{\nolinebreak\hspace{0.1em}]},  {\em
  Society of Photo-Optical Instrumentation Engineers (SPIE) Conference Series}
  {\bf 11451},  1145108 (Dec. 2020).

\bibitem{Pike2020}
{Pike}, F.~A., {Sharma}, T.~K., {Beno{\^\i}t}, A., {MacLachlan}, D.~G.,
  {Dinkelaker}, A.~N., {Nayak}, A.~S., {Madhav}, K., {Roth}, M.~M., {Labadie},
  L., {Pedretti}, E., {ten Brummelaar}, T.~A., {Scott}, N.~J., {Coud{\'e} du
  Foresto}, V., and {Thomson}, R.~R., ``{K-band integrated optics beam
  combiners for CHARA fabricated by ultrafast laser inscription},'' in [{\em
  Society of Photo-Optical Instrumentation Engineers (SPIE) Conference
  Series}{\nolinebreak\hspace{0.1em}]},  {\em Society of Photo-Optical
  Instrumentation Engineers (SPIE) Conference Series} {\bf 11446},  114460K
  (Dec. 2020).

\bibitem{Siliprandi2022}
{Siliprandi}, J., {MacLachlan}, D.~G., {Ross}, C.~A., {Sharma}, T.~K.,
  {Labadie}, L., {Madhav}, K., {Nayak}, A.~S., {Dinkelaker}, A.~N., {Roth},
  M.~M., {Pedretti}, E., {ten Brummelaar}, T.~A., {Scott}, N.~J., {Coud{\'e} du
  Foresto}, V., {Thomson}, R.~R., and {Beno{\^\i}t}, A., ``{Ultrafast laser
  inscription of integrated optics two-telescope beam combiners for K-band
  interferometry at the CHARA array},'' in [{\em Optical and Infrared
  Interferometry and Imaging VIII}{\nolinebreak\hspace{0.1em}]},  {M{\'e}rand},
  A., {Sallum}, S., and {Sanchez-Bermudez}, J., eds., {\em Society of
  Photo-Optical Instrumentation Engineers (SPIE) Conference Series} {\bf
  12183},  1218314 (Aug. 2022).

\bibitem{Bland-Hawthorn2011}
{Bland-Hawthorn}, J., {Ellis}, S.~C., {Leon-Saval}, S.~G., {Haynes}, R.,
  {Roth}, M.~M., {L{\"o}hmannsr{\"o}ben}, H.~G., {Horton}, A.~J., {Cuby},
  J.~G., {Birks}, T.~A., {Lawrence}, J.~S., {Gillingham}, P., {Ryder}, S.~D.,
  and {Trinh}, C., ``{A complex multi-notch astronomical filter to suppress the
  bright infrared sky},'' {\em Nature Communications}~{\bf 2},  581 (Dec.
  2011).

\bibitem{BlandHawthorn2016}
{Bland-Hawthorn}, J., {Min}, S.-S., {Lindley}, E., {Leon-Saval}, S., {Ellis},
  S., {Lawrence}, J., {Beyrand}, N., {Roth}, M., {L{\"o}hmannsr{\"o}ben},
  H.-G., and {Veilleux}, S., ``{Multicore fibre technology: the road to
  multimode photonics},'' in [{\em Advances in Optical and Mechanical
  Technologies for Telescopes and Instrumentation
  II}{\nolinebreak\hspace{0.1em}]},  {Navarro}, R. and {Burge}, J.~H., eds.,
  {\em Society of Photo-Optical Instrumentation Engineers (SPIE) Conference
  Series} {\bf 9912},  99121O (July 2016).

\bibitem{Trinh2012}
{Trinh}, C.~Q., {Ellis}, S.~C., {Lawrence}, J.~S., {Horton}, A.~J.,
  {Bland-Hawthorn}, J., {Leon-Saval}, S.~G., {Bryant}, J., {Case}, S.,
  {Colless}, M., {Couch}, W., {Freeman}, K., {Gers}, L., {Glazebrook}, K.,
  {Haynes}, R., {Lee}, S., {L{\"o}hmannsr{\"o}ben}, H.~G., {Miziarski}, S.,
  {O'Byrne}, J., {Rambold}, W., {Roth}, M.~M., {Schmidt}, B., {Shortridge}, K.,
  {Smedley}, S., {Tinney}, C.~G., {Xavier}, P., and {Zheng}, J., ``{GNOSIS: a
  novel near-infrared OH suppression unit at the AAT},'' in [{\em Ground-based
  and Airborne Instrumentation for Astronomy IV}{\nolinebreak\hspace{0.1em}]},
  {McLean}, I.~S., {Ramsay}, S.~K., and {Takami}, H., eds., {\em Society of
  Photo-Optical Instrumentation Engineers (SPIE) Conference Series} {\bf 8446},
   84463J (Sept. 2012).

\bibitem{Trinh2013}
{Trinh}, C.~Q., {Ellis}, S.~C., {Bland-Hawthorn}, J., {Lawrence}, J.~S.,
  {Horton}, A.~J., {Leon-Saval}, S.~G., {Shortridge}, K., {Bryant}, J., {Case},
  S., {Colless}, M., {Couch}, W., {Freeman}, K., {L{\"o}hmannsr{\"o}ben},
  H.-G., {Gers}, L., {Glazebrook}, K., {Haynes}, R., {Lee}, S., {O'Byrne}, J.,
  {Miziarski}, S., {Roth}, M.~M., {Schmidt}, B., {Tinney}, C.~G., and {Zheng},
  J., ``{GNOSIS: The First Instrument to Use Fiber Bragg Gratings for OH
  Suppression},'' {\em AJ}~{\bf 145},  51 (Feb. 2013).

\bibitem{Ellis2018}
{Ellis}, S.~C., {Bauer}, S., {Bacigalupo}, C., {Bland-Hawthorn}, J., {Bryant},
  J.~J., {Case}, S., {Content}, R., {Fechner}, T., {Giannone}, D., {Haynes},
  R., {Hernandez}, E., {Horton}, A.~J., {Klauser}, U., {Lawrence}, J.~S.,
  {Leon-Saval}, S.~G., {Lindley}, E., {L{\"o}hmannsr{\"o}ben}, H.~G., {Min},
  S.~S., {Pai}, N., {Roth}, M., {Shortridge}, K., {Waller}, L., {Xavier}, P.,
  and {Zhelem}, R., ``{PRAXIS: an OH suppression optimised near infrared
  spectrograph},'' in [{\em Ground-based and Airborne Instrumentation for
  Astronomy VII}{\nolinebreak\hspace{0.1em}]},  {Evans}, C.~J., {Simard}, L.,
  and {Takami}, H., eds., {\em Society of Photo-Optical Instrumentation
  Engineers (SPIE) Conference Series} {\bf 10702},  107020P (July 2018).

\bibitem{Ellis2020}
{Ellis}, S.~C., {Bland-Hawthorn}, J., {Lawrence}, J.~S., {Horton}, A.~J.,
  {Content}, R., {Roth}, M.~M., {Pai}, N., {Zhelem}, R., {Case}, S.,
  {Hernandez}, E., {Leon-Saval}, S.~G., {Haynes}, R., {Min}, S.~S., {Giannone},
  D., {Madhav}, K., {Rahman}, A., {Betters}, C., {Haynes}, D., {Couch}, W.,
  {Kewley}, L.~J., {McDermid}, R., {Spitler}, L., {Sharp}, R.~G., and
  {Veilleux}, S., ``{First demonstration of OH suppression in a high-efficiency
  near-infrared spectrograph},'' {\em MNRAS}~{\bf 492},  2796--2806 (Feb.
  2020).

\bibitem{Ellis2020a}
{Ellis}, S.~C., {Bland-Hawthorn}, J., {Bauer}, S., {Case}, S., {Content}, R.,
  {Fechner}, T., {Giannone}, D., {Haynes}, R., {Hernandez}, E., {Horton},
  A.~J., {Klauser}, U., {Lawrence}, J.~S., {Leon-Saval}, S.~G.,
  {L{\"o}hmannsr{\"o}ben}, H.~G., {Min}, S.~S., {Pai}, N., {Roth}, M.,
  {Waller}, L., and {Zhelem}, R., ``{PRAXIS: an OH suppression optimised near
  infrared spectrograph},'' in [{\em Advances in Optical Astronomical
  Instrumentation 2019}{\nolinebreak\hspace{0.1em}]},  {Ellis}, S.~C. and
  {d'Orgeville}, C., eds., {\em Society of Photo-Optical Instrumentation
  Engineers (SPIE) Conference Series} {\bf 11203},  1120312 (Jan. 2020).

\bibitem{Goebel2018}
{Goebel}, T.~A., {Bharathan}, G., {Ams}, M., {Richter}, D., {Kr{\"a}mer},
  R.~G., {Heck}, M., {Siems}, M.~P., {Fuerbach}, A., and {Nolte}, S.,
  ``{Ultrashort pulse point-by-point written aperiodic fiber Bragg gratings for
  suppression of OH-emission lines},'' in [{\em Advances in Optical and
  Mechanical Technologies for Telescopes and Instrumentation
  III}{\nolinebreak\hspace{0.1em}]},  {Navarro}, R. and {Geyl}, R., eds., {\em
  Society of Photo-Optical Instrumentation Engineers (SPIE) Conference Series}
  {\bf 10706},  107066N (July 2018).

\bibitem{Rahman2020}
{Rahman}, A., {Madhav}, K., and {Roth}, M.~M., ``{Complex phase masks for OH
  suppression filters in astronomy: part I: design},'' {\em Optics
  Express}~{\bf 28},  27797 (Sept. 2020).

\end{thebibliography}
\bibliographystyle{spiebib} 

\end{document}